\documentclass[useAMS,usenatbib]{mn2e}
\usepackage{epsfig}
\usepackage{graphics}
\newcommand{\aaps}{\rm A$\&$AS}

\newcommand{\Ha}{H$\alpha$}

\newcommand{\Hg}{H$\gamma$}
\newcommand{\Hd}{H$\delta$}

\newcommand{\kms}{ km\ s$^{-1}$}                            
\newcommand{\Appx}{ \AA\ px$^{-1}$}             

 \title[A method for the Determination of Stellar Population Parameters]
{An Active Instance-based Machine Learning method for Stellar Population
Studies}
\author[T. Solorio et al.]{Thamar Solorio$^{1}$\thanks{E-mail: thamy@inaoep.mx(TS); fuentes@inaoep.mx(OF); rjt@inaoep.mx(RT);
eterlevi@inaoep.mx(ET)} Olac Fuentes$^{1}$ \footnotemark[1] Roberto Terlevich$^{1,2}$\footnotemark[1] and Elena Terlevich$^{1}$\footnotemark[1]\\
$^{1}$Instituto Nacional de Astrof\'\i sica, \'Optica y
Electr\'onica, INAOE,  Luis Enrique Erro 1. Tonantzintla, Puebla
72840, M\'exico \\
$^{2}$ Institute of Astronomy, Madingley Road, Cambridge CB3 0HA,
United Kingdom}
\begin{document}

\date{Accepted. Received ; in original form ;}
\pagerange{\pageref{firstpage}--\pageref{lastpage}} \pubyear{2004}

\maketitle

\label{firstpage}

\begin{abstract}
We have developed a method for fast and accurate stellar population parameters
determination in order to apply it to high resolution galaxy spectra. 
The method is based on an
optimization technique that combines active learning with an instance-based
machine learning algorithm.

We tested the method with the retrieval of the star-formation
history and dust content in ``synthetic" galaxies with a wide
range of S/N ratios. The ``synthetic" galaxies where constructed
using two different grids of high resolution theoretical
population synthesis models.

The results of our controlled experiment shows that our method can estimate with
good speed and accuracy the parameters of the stellar
populations that make up the galaxy even for very low S/N input.
For a spectrum with S/N$=5$ the typical average deviation between the input and
fitted spectrum is
less than  $10^{-5}$. Additional improvements are achieved using prior knowledge.

\end{abstract}


\section{Introduction}

The availability of large astronomical spectroscopic surveys with
moderate spectral resolution such as the 2dF (Colless et al.~ 2001) or the Sloan
Digital Sky Survey (SDSS, York et al.~2000; Stoughton et al.~2002), has prompted
the computation of new grids of high resolution spectral synthesis
models creating the need of highly efficient methods for the
determination of intrinsic physical parameters of a large number
of galaxies. There are three intrinsic galactic parameters that are
particularly important for studies of cosmological evolution:
The star formation and chemical composition histories and the mass
distribution of their stellar populations. The importance of the
accurate knowledge of these parameters for cosmological studies
and for the understanding of galaxy formation and evolution cannot
be overestimated. Template fitting has been widely used to carry
out estimates of the distribution of age and metallicity from
spectral data. Although this technique achieves good results, it
is expensive in terms of computing time  (therefore is best
applied to relatively small samples e.g.~Mayya et al. 2004)
and the results are in general compromised by the low signal-to-noise
data (Kauffmann et al.~2003; Tremonti et al.~2004; Cid fernandes et al. 2005).

Until recently, synthesis models provided either low resolution and a range of metallicities
using theoretical atmospheres, or medium resolution at basically solar abundance with the
use of empirical stellar spectra. A major problem with theoretical atmospheres used to be
that the sampling was coarser than the line broadening observed even in the most
massive galaxies. For massive ellipticals with velocity dispersion of up
to 400\kms sampling about or better than 7\Appx\ in the
optical region is needed for representing their spectra with minimum loss of
information. For dwarf galaxies or globular clusters with velocity dispersion all the way
down to 5-10\kms the optimum sampling is around 0.1 \Appx.

Clearly comparing data obtained with sampling of 70\kms\ like the
SDSS with models with sampling of 1200\kms\ at 5000\AA\   is not
satisfactory in the sense that much information associated
with atomic lines and even relatively narrow molecular bands will
be washed out by the large under sampling. On the other hand, by
smoothing or filtering the high frequencies in the data, a more
compact and easier/faster to process data set is created (Heavens,
Jim\'enez and Lahav 2000, Heavens et al. 2004).

To overcome these problems we have explored new methods that,
while exploiting the high resolution achieved by recent synthesis
models, maximize both speed and accuracy in the determination of
stellar population parameters.

Minimum distance methods and the closely related chi-square minimization
present two significant drawbacks as classification tools. Firstly,
they depend crucially on the choice of standard objects to be used;
in many problems, it is impossible to select a representative member
for each class or each combination of parameters. Secondly, it is
difficult to include information regarding intra-class variability,
as the only information provided is a typical representative member of
the class. Machine learning approaches, on the other hand, use training
data that include many representative examples for each class, which
makes the selection of standards unnecessary and provides information
to determine the features that discriminate members of different classes.

Machine learning algorithms have been shown to significantly 
outperform minimum distance methods and chi-square minimization in
a number of astronomical applications, including stellar
classification and determination of stellar atmospheric
parameters. For instance, Bailer-Jones (1996) showed that a
committee of simple feedforward neural networks yields an error
reduction of about 50 percent compared with minimum distance when
applied to the task of stellar classification; similar results
were reported in Gulati and Gupta (1997).
Bailer-Jones (1996) also mentions the
fact that, for regions where training data are sparse, the performance
advantage of neural network decreases. Thus, methods that automatically
add training data to undersampled regions, as the one we present in this 
paper, are highly desirable.

In this, the first paper of a series, we test a technique that
approximates non-linear multidimensional functions using a small
initial training set, and by using active learning it increases
this training set as needed according to the elements of the test
set. This method has shown to outperform traditional instance-base
learning algorithms on the problem of interferogram analysis
(Fuentes and Solorio, 2004).

Here we present the results of a series of controlled experiments
showing that this method can quickly and accurately retrieve the
physical parameters of ``simulated" galaxies, even at a very low
S/N level. Our method takes also advantage of prior domain
knowledge which is used to further increase the accuracy of the
results obtained. 
In a forthcoming paper (Solorio et al., in preparation)
we apply this methodology to large data sets of
galaxy spectra to characterize their stellar population fabric.

\section{Testing the method with Synthetic galaxies}

Before blindly applying a new method to real data it is reasonable
to critically test the procedure in a controlled environment.
A crucial aspect is that the
validity of the test increases as the test conditions approach
the real case. For this reason we have created synthetic galaxies
as realistic as possible and necessary in this first step in our
research. We thus have applied our methods to a reference set of
``synthetic'' high resolution spectra of galaxies. To minimize
systematics associated with the use of a particular model we have
used two different sets of new high resolution spectral synthesis
models (for this test only solar metallicity ones) to generate the
reference synthetic galaxy spectra set (Bertone et al.~2004:
Padova models; Gonz\'alez-Delgado et al.~2005: Granada models).
The high spectral resolution of the models, allows to use them in
the study of narrow absorption lines and for the spectral
evolution of the intense line profiles over a wide range of ages.
It should be emphasized that our goal is to test the effectiveness
of the method in two different sets of models, in order to assess
its robustness. 
We are not trying to determine the respective
merits of the models, thus our experiments do not give evidence of
any of this. We will address this point in a forthcoming paper.

The Granada models are Single Stellar Population (SSP) synthesis
calculated for ages ranging from 1~Myr to 17~Gyr using the Padova and
Geneva stellar evolutionary tracks, and their own stellar
atmospheres library with spectral sampling of 0.3 \AA , and a
wavelength coverage of 3000-7000 \AA\ (Martins et al.~2005) .
Of the various models
available regarding metallicities, we only use for this first work
the solar metallicity ones. The synthetic stellar library has been
computed with the latest stellar atmospheres, non-LTE for the hot
and LTE line-blanketed models for the cold stars. A full
description of the models is given by Gonz\'alez-Delgado et al.~2005.

A second set of integrated high resolution spectra that we will call the
Padova set, was kindly computed for us
by A. Bressan (private communication) according to the
prescriptions outlined in Bressan, Chiosi and Fagotto (1994).
Spectral fluxes along the Bertelli et al. (1994) isochrones were
integrated adopting a Salpeter initial mass function (IMF) between
0.15 and 120 $M_{\odot}$. Kurucz high resolution (R=50000)
synthetic stellar spectra from 3500\AA\ to 4500\AA, were kindly
provided by L. Rodr\'\i guez, M. Ch\'avez and E. Bertone before
publication (Rodr\'\i guez-Merino et al.~2005). The red end of the spectra
was completed using their 20 \AA\ resolution models from 4500 to 7000 \AA\
(Bressan et al.~1994).
The spectral resolution of the SSPs where finally degraded to R=10000.

\subsection{Synthetic galaxies}\label{s:synthetic}

To construct the spectrum of the synthetic galaxies we combined
three different populations corresponding to young, intermediate
age and old single-age stellar populations (SSP) in varying proportions.
To each population we added independent dust attenuation (extinction).
The effects of adding noise are discussed in the next section.

Let $f(\lambda)$ be the energy flux emitted by a star or group of
stars at wavelength $\lambda$. The flux detected by a measuring
device is then $d(\lambda) = f(\lambda) (1-e^{-r\lambda})$, where
$r$ is a constant that defines the amount of reddening in the
observed spectrum and depends on the size and density of the dust
particles in the interstellar medium.

A synthetic galactic spectrum, $g(\lambda)$, can be built given
$c_1,c_2,c_3$, the relative contributions of young, intermediate
age and old stellar populations, respectively, their reddening
parameters $r_1,r_2,r_3$, and the ages of the populations
$a_1,a_2,a_3$.
\begin{equation}\label{g}
g(\lambda) = \sum_{i=1}^3 c_i s(a_i,\lambda) (1-e^{-r_i\lambda})
\end{equation}
 where $g(\lambda)$ is the energy flux
detected at wavelength $\lambda$ and $s(a_{i},\lambda)$ is the
flux emitted by a stellar population of age $a_{i}$ at wavelength
$\lambda$.

The task of analyzing an observed galaxy spectrum $t$ consists of
finding the parameter vector $q =
[c_1,c_2,c_3,r_1,r_2,r_3,a_1,a_2,a_3]$ that minimizeS:
\begin{equation}\label{eq_error}
{\it error} (q) = \sum_\lambda (t(\lambda) - g(\lambda))^{2}
\end{equation}

Clearly, $c_1,...,c_3$ have to be non-negative, and sum up to 1,
also, realistic values of $r_1,...,r_3$ are in the narrow range
$[1\times10^{-5}, 6\times10^{-4}]$, and using only a few discrete
values for $a_1,a_2$ and $a_3$ normally suffices for a reasonable
approximation. In particular, for our experiments we consider stellar
population ages $a_1\in \{3\times10^{6}\}$, $a_2 \in \{10^{8}, 3\times10^{8},
5\times10^{8}, 8\times10^{8}\}$, and $a_3 \in \{10^{9},
2\times10^{9}, 3\times10^{9} ,5\times 10^{9}, 10^{10}\}$.

\section{The Method}\label{s:method}

\begin{table}
\hrule
\begin{tabbing}
\\
 12\=12\=12\= 12\=\kill
  0. Let $T$ be the test spectra \\
  1. $S = \{\}$\\
  2. For $i=1$ to n \\
\>    2.1. Generate random parameter vector $p=[c_1,c_2,c_3,r_1,r_2,r_3,$\\
\hspace{.6cm} $a_1,a_2,a_3]$ \\
\>  2.2. Generate spectra $s$ according to $p$ \\
\>  2.3. $ S = S \cup \{\langle \langle s \rangle, \langle
r_1,r_2,r_3\rangle\rangle$\}\\
3. While $T \neq \{\}$ do: \\
\> 3.1. Build $C$, an ensemble of approximators using learning \\
\> \> algorithm LWLR and training data $S$\\
\> 3.2. For every test spectra $t \in T$ \\
\> \> 3.2.1. Use $C$ to predict the reddening parameters $r_1,r_2,r_3$ of $t$ \\
\> \> 3.2.2. {\it error}($q^*$) = $\infty$\\
\> \> 3.2.3. For every triple $\langle a_1,a_2,a_3 \rangle \in \{3\times10^{6}\} \times \{10^{8}, 3\times10^{8},$\\
\hspace{.8cm} $ 5\times10^{8}, 8\times10^{8}\} \times$ $\{10^{9}, 2\times10^{9}, 3\times10^{9} ,5\times 10^{9}, 10^{10}\}$ \\
\> \> \> $ R = \left[ \begin{array}{c} s(a_1,\lambda_1)
(1-e^{-r_1\lambda_1}), ..., s(a_1,\lambda_m)
(1-e^{-r_1\lambda_m}) \\
s(a_2,\lambda_1) (1-e^{-r_2\lambda_1}), ..., s(a_2,\lambda_m)
(1-e^{-r_2\lambda_m}) \\s(a_3,\lambda_1) (1-e^{-r_3\lambda_1}),
..., s(a_3,\lambda_m)
(1-e^{-r_3\lambda_m}) \\
\end{array} \right]$ \\
\>\>\>$ [c_1,c_2,c_3] = t (R^T R)^{-1} R^T$ \\
\>\>\> Generate spectra $g$ according to $q=[c_1,c_2,c_3,r_1,r_2,r_3,$\\
\hspace{.8cm} $a_1,a_2,a_3]$ \\
\>\>\> {\it error}($q$) = $\sum_\lambda (g(\lambda) - t(\lambda))^{2}$\\
\>\>\>  If {\it error}($q) < $  {\it error} $(q^*)$\\
\>\>\> \> $q^* = q$\\
\>\>\> \> $g^* = g$\\
\>\> 3.2.4. If {\it error}$(q^*) < ${\it threshold}\\
\>\>\> output $\langle t, q^*\rangle $ \\
\>\>\> $T = T - \{t\}$\\
\>\> \> Else $S = S \cup \{\langle \langle g^* \rangle, \langle
r_1,r_2,r_3\rangle\rangle$\}\\
\end{tabbing}
\hrule 
 \caption{Pseudo-code of our Active Learning
Algorithm (described in Section~\ref{s:method}).}
 \label{t:pseudo}
\end{table}

In the application proposed  here, galactic spectral analysis, the
algorithm estimates the ages of three SSP, 
their individual
contribution to the total light plus the reddening from a high
resolution or equivalently high dimensionality input spectrum. In
general, all learning algorithms, such as neural networks, C4.5
(Quinlan 1993), and locally weighted regression, face the
well known curse of dimensionality (Bellman 1957), which
essentially states that the number of training examples needed to
approximate a function accurately grows exponentially with the
dimensionality of the task. To circumvent the curse of
dimensionality, we partition the problem into three subproblems,
each of which is amenable to be solved by a different method. The
key point is that the dust extinction is a non-linear effect that
takes long to estimate, thus if the values of the reddening
parameters were known, it would be possible to just perform a
search over the possible combinations of values for the ages of
stellar populations (a total of $1 \times 4 \times 5$ = 20 for the
Granada models, and a total of 16 for the Padova models), and for
each combination of ages find the contributions that best fit the
observation using least squares. Then the best overall fit would
be the combination of ages and contributions that resulted in the
best match to the test spectrum. Thus, the crucial sub-problem to
be solved is that of determining the reddening parameters.

Predicting the reddening parameters from spectra is a difficult
non-linear optimization problem, specially for the case of noisy
spectra. We propose to solve it using an iterative active learning
algorithm that learns the function from spectra to reddening
parameters. In each iteration, the algorithm uses its training set
to build an approximator to predict the reddening parameters of
the spectra in the test set. Once the algorithm has predicted
these parameters, it uses them to find the combination of ages and
contributions that yield the best match to the observed spectra.
>From these parameters we can generate the corresponding spectrum,
and compare it with the spectrum under analysis, if they are a
close match, then the parameters found by the algorithm are
correct, if not, we can add the newly generated training example
(the predicted parameters and their corresponding spectrum) to the
training set and proceed to a new iteration. Since this type of
active learning adds to the training set examples that are
progressively closer to the points of interest, the errors are
guaranteed to decrease in every iteration until convergence is
attained. In this algorithm the criteria to halt the iterative
process can be an error threshold, or a maximum number of
iterative steps.

An outline, in the form of pseudocode, of the algorithm is given
in table~\ref{t:pseudo}. In steps 1 and 2 we build an initial
training set $S$ containing $N$ spectra (the attributes),
generated for randomly chosen parameter vectors (the target
function), applying equation \ref{g}. Step 3 forms the main loop,
in which we will attempt to obtain the parameters that best match
the spectra under analysis (set $T$), this step is repeated until
a satisfactory fit has been found for every spectrum in the test
set. First, in step 3.1, an approximator $C$ is built using $S$
and an ensemble and locally weighted linear regression (LWLR), a
well-known instance-based learning algorithm (Atkeson et al.~1997).
Using $C$, we obtain candidate reddening parameters
$[r_1,r_2,r_3]$ for each spectrum in the test set $T$, this is the
non-linear part of the problem (step 3.2.1). Given the candidate
reddening parameters, in step 3.2.3 we find the ages of the
stellar populations $[a_1,a_2,a_3]$ and their relative
contributions $[c_1,c_2,c_3]$ using a combination of exhaustive
search and least squares fitting. For each of the possible 20
combinations of ages we find the relative contributions that best
match the spectrum under analysis using a pseudo-inverse
computation and then choose among the 20 combination of ages and
corresponding contributions the one that minimizes the residuals
(equation \ref{eq_error}). In step 3.2.4 we simply test if the
parameter vector results in a satisfactory fit, if the error for
the best approximation, computed as depicted in
equation~\ref{eq_error}, is smaller than a set threshold, it
outputs the set of parameters found for that spectrum and removes
it from the test set, if the error is not small enough, it adds
the new training example to the training set and continues the
process.

It should be pointed out that the active learning algorithm is
independent of the choice of base learning algorithm used to
predict the reddening parameters. Any algorithm that is suitable
to predict real-valued target functions from real-valued
attributes could be used. In this work we use an ensemble of
locally-weighted linear regression (LWLR), 
but others such as
K-nearest-neighbours could have been applied. In the following two
sections we briefly present the ideas behind our chosen base
learning algorithm.

\subsection{Ensembles}
An ensemble of classifiers is a set of classifiers whose
individual decisions are combined in some way, normally by voting.
In order for an ensemble to work properly, individual members of
the ensemble need to have uncorrelated errors and an accuracy
higher than random guessing. There are several methods for
building ensembles. One of them, which is called \textit{bagging}
(Breiman 1996), consists of manipulating the training set. In this
technique, each member of the ensemble has a training set
consisting of \textit{m} examples selected randomly with
replacement from the original training set of \textit{m} examples
(Dietterich 2000). Another technique similar to bagging
manipulates the attribute set. Here, each member of the ensemble
uses a different subset randomly chosen from the attribute set.
More information concerning ensemble methods, such as boosting and
error-correcting output coding, can be found in (Dietterich 2000).
The technique used for building an ensemble is chosen according to
the learning algorithm used, which in turn is determined by the
learning task. In the work presented here, we use the technique
that randomly selects subsets of attributes.

\subsection{Locally-Weighted Regression}\label{s:LWLR}

Locally-Weighted Regression (LWR) belongs to the family of
instance-based learning algorithms, which includes algorithms as
the basic K-nearest neighbour and radial basis functions (Powell
1987). In contrast to most other learning algorithms, which use
their training examples to construct explicit global
representations of the target function, instance-based learning
algorithms simply store some or all of the training examples and
postpone any generalization effort until a new instance must be
classified. They can thus build query-specific local models, which
attempt to fit the training examples only in a region around the
query point. In this work we use a linear model around the query
point to approximate the target function.

Given a query point ${\bf x_q}$, to predict its output parameters
${\bf y_q}$, we find the $k$ examples in the training set that are
closest to it, and assign to each of them a weight given by the
inverse of its distance to the query point: $w_i = \frac{1}{| {\bf
x_q} - {\bf x_i}|}$. Let $W$, the weight matrix, be a diagonal
matrix with entries $w_1,\dots,w_n$. Let $X$ be a matrix whose
rows are the vectors ${\bf x_1}, \dots, {\bf x_k}$, the input
parameters of the examples in the training set that are closest to
${\bf x_q}$, with the addition of a  ``1'' in the last column. Let
$Y$ be a matrix whose rows are the vectors ${\bf y_1}, \dots, {\bf
y_k}$, the output parameters of these examples. Then the weighted
training data are given by $Z = W X$ and the weighted target
function is $V = W Y$. Then we use the estimator for the target
function ${\bf y_q} = {\bf x_q}^T (Z^T Z) ^{-1} Z^T V.$

Thus, locally weighted linear regression is very similar to
least-squares linear regression, except that the error terms used
to derive the best linear approximation are weighted by the
inverse of their distance to the query point. Intuitively, this
yields much more accurate results than standard linear regression
because the assumption that the target function is linear does not
hold in general, but is a good approximation when only a small
size neighborhood is considered.

\section{Discussion}In all the experiments reported
here we used the following procedure: firstly we generated a
random set of 200 galactic spectra with their corresponding
parameters. This set was then randomly divided into two disjoint
subsets, one subset consisting of 50 galactic spectra was used for
training and the remaining 150 was considered the test set. This
procedure was repeated 10 times, and we report here the overall
mean results.

\begin{figure}
\centering
 \includegraphics[height=65mm, width=65mm]{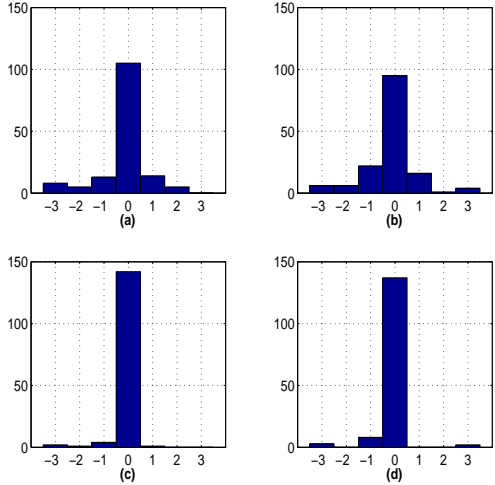}\\
 \caption{Distribution of errors in the age prediction of intermediate and old populations using the Granada models.
 Error in age prediction is measured as the distance in logarithmic steps between the age of the test spectrum and the predicted age.
 Figure (a), intermediate age, and (b), old,
are the predictions of a traditional LWLR ensemble.
 Figures (c) and (d) are the predictions of our algorithm for the same ages and test spectra.} \label{f:histGranada}
\end{figure}

\begin{figure}
\centering
 \includegraphics[height=75mm, width=85mm]{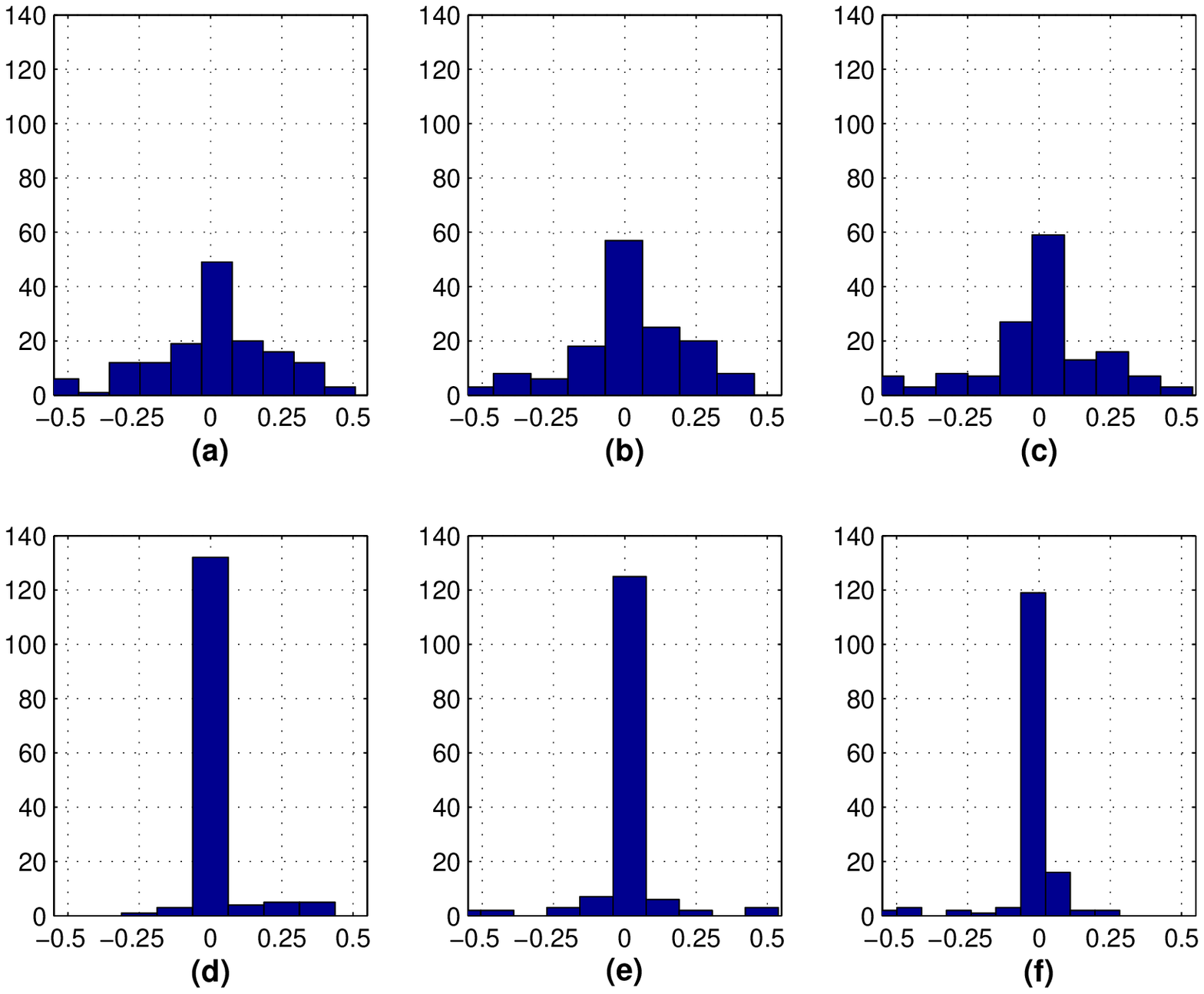}\\
 \caption{Distribution of prediction errors in the relative contribution parameters $c_{1}$,$c_{2}$, and $c_{3}$
 (see section \ref{s:synthetic}), using the Granada models. Figures (a) to (c) are the predictions using an
 ensemble of LWLR for young, intermediate and old populations; figures (d) to (f) are predictions using our active learning
algorithm.}
  \label{f:hContGranSinRuido}
\end{figure}

\begin{figure}
\centering
 \includegraphics[height=75mm, width=85mm]{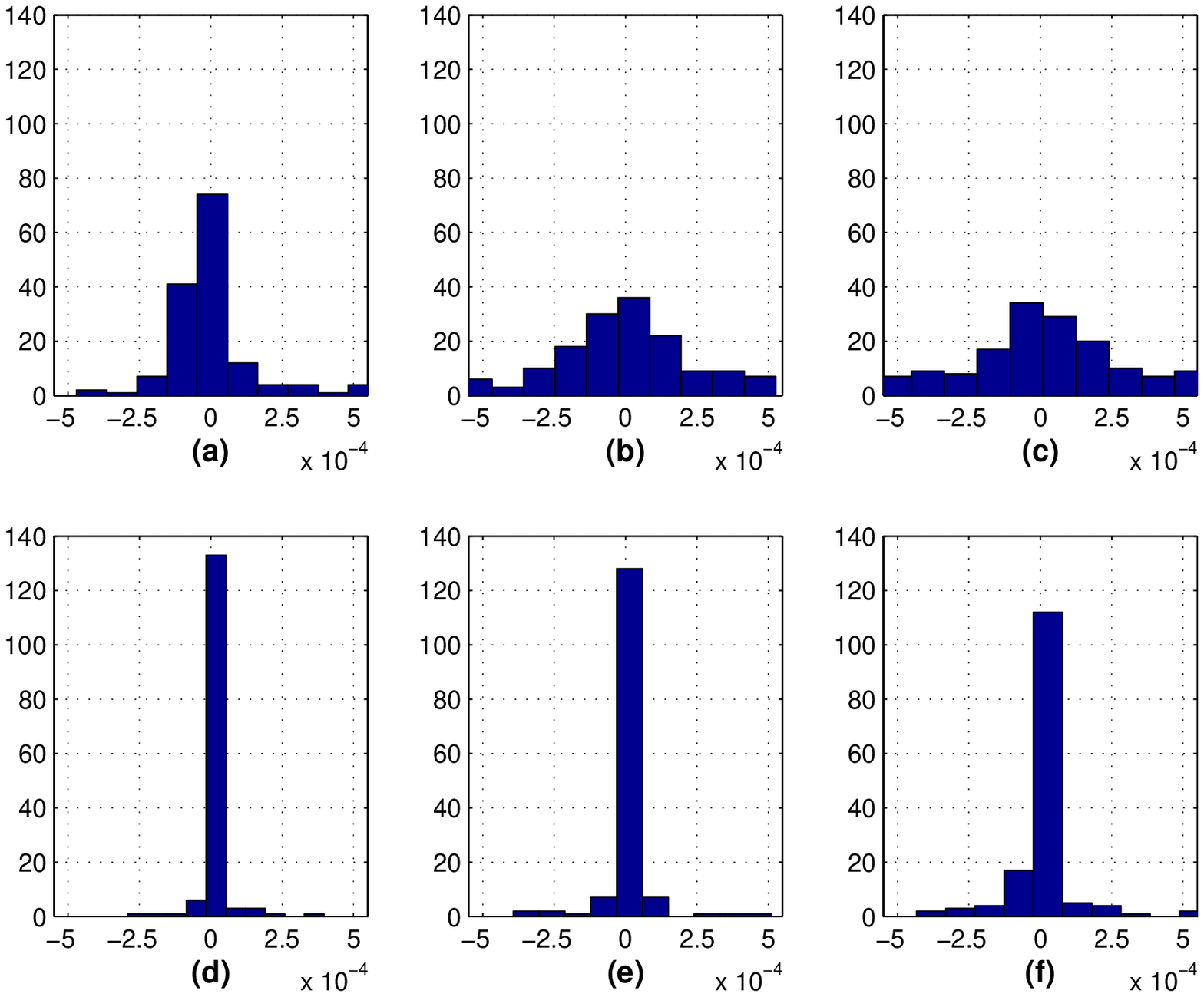}\\
 \caption{Distribution of prediction errors in the reddening parameters $r_{1}$, $r_{2}$ and $r_{3}$, (see section
 \ref{s:synthetic}), using the Granada models. Figures (a) to (c) are the predictions for young, intermediate and old populations using an ensemble of LWLR, figures (d) to (f) are predictions using our active learning algorithm. }
  \label{f:hRedGranSinRuido}
\end{figure}

In the first set of experiments our objective was to determine
empirically the differences between the active learning procedure
versus a traditional ensemble of LWLR. As mentioned previously,
the ensembles were constructed selecting randomly a subset of the
attributes. To make the comparison objective, both methods used
the same attribute subset and an ensemble of size 5. In
Figure~\ref{f:histGranada} we show the distribution for prediction
errors in intermediate and old ages using the Granada models. This
error is measured as the distance in logarithmic steps between the
real age and the predicted one. We can see that even though the
traditional ensemble of LWLR performs well, our active
algorithm achieves higher accuracy.
Figure~\ref{f:hContGranSinRuido} shows error distributions
corresponding to the prediction of relative contributions:
$c_{1}$,$c_{2}$ and $c_{3}$, of each age population for the
Granada models also. We can see that for the active algorithm the
central bars are higher than those of an ensemble of LWLR.
Error distributions for prediction of reddening parameters are shown in
Figure~\ref{f:hRedGranSinRuido}. Comparable results in all the
experiments were obtained using the Padova models, in
Tables~\ref{t:AgePado}, \ref{t:ContPado} and \ref{t:RedPado} we
present the more discrepant results.

\begin{figure}
\centering
 \includegraphics[height=105mm, width=85mm]{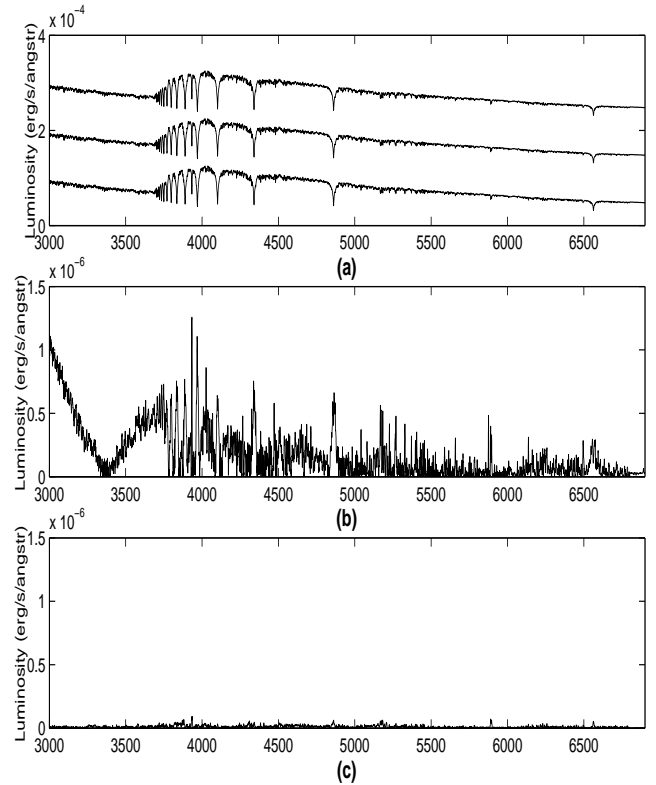}\\
 \caption{Graphical comparison of results using the Granada models. Figure (a) from top to bottom
 and shifted by a constant to aid visualization: original test spectrum,
 spectrum recovered using ensemble of LWLR and spectrum recovered using active learning. Figures (b) and (c) show, in the same
 scale, the residuals between test and predicted spectra in the same listed order.}
  \label{f:granSinRuido}
\end{figure}

\begin{figure}
\centering
 \includegraphics[height=105mm, width=85mm]{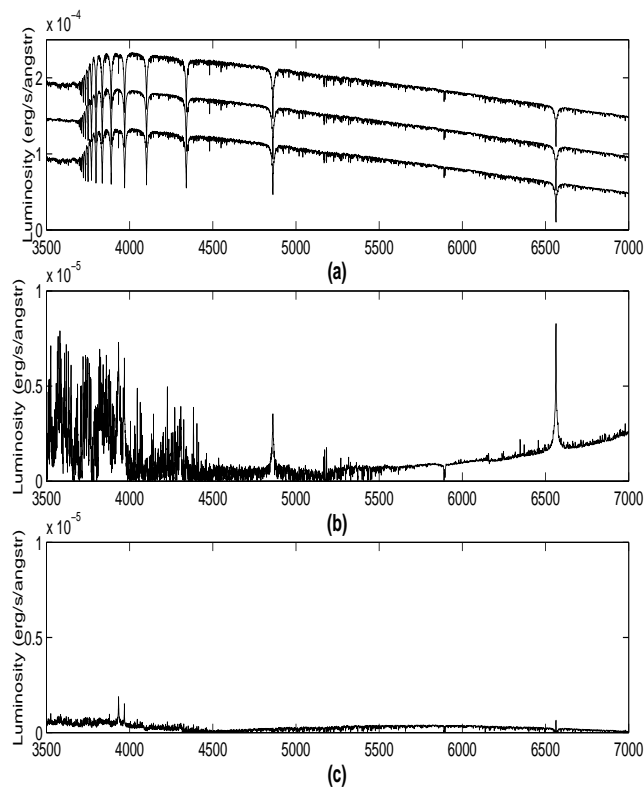}\\
 \caption{Same as Figure~\ref{f:granSinRuido} but using the Padova models. Figure (a) from top to bottom
 and shifted by a constant to aid visualization: original test spectrum,
 spectrum recovered using ensemble of LWLR and spectrum recovered using active learning. Figures (b) and (c) show, in the same scale, the
 residuals between test spectrum and predicted spectra in the same listed
order. The red end of the spectra
was completed using Padova 20 \AA\ resolution models from 4500 to 7000 \AA\
(Bressan et al.~1994).}
  \label{f:padSinRuido}
\end{figure}

Figures~\ref{f:granSinRuido} and \ref{f:padSinRuido} show
graphical comparisons between a test spectrum, a reconstructed
spectrum using traditional LWLR and our active learning technique
for both models. The residuals are always smaller than 3 percent
for the LWLR method and clearly much smaller for the active
learning technique. 
The fact that the active learning technique
outperformed the traditional ensemble of LWLR was not surprising. 
Although both are based on the same learning algorithm, LWLR, the 
training sets from which the predictions are computed are different.
The main difference between these two techniques lies on the iterative
process of the active learning algorithm. In each iteration, the active 
learning algorithm augments its training set with new examples that will
allow it to better approximate the observed spectra. And this iterative
process continues until a suitable solution is found for each spectrum
in the test set. As the traditional ensemble of LWLR lacks this iterative
process, it will output the best predictions it can reach using only the
original training set. 

\subsection{The effect of noise}
The results presented above are very encouraging. However, the data used in
those experiments were noise free. We are aware that noisy data
pose a more realistic evaluation of our algorithm, given that in
real data analysis noise is always present. Astronomical spectral
analysis is no exception to this rule. For this reason, we have
performed a set of experiments aimed at exploring the
noise-sensitivity of our active learning algorithm. We performed
the same procedure described previously, except that this time we
added to the test data a Gaussian noise with zero mean and
standard deviation of one. We experimented with three different
signal to noise (S/N) ratios: 5, 30 and 100 corresponding to bad,
normal and good data respectively. Here we present only results
for the lower S/N level, given that as the noise level decreases
error predictions are more similar to noiseless data experiments.

As a first stage in the treatment of noisy data we used a
procedure involving standard principal component analysis (PCA).
PCA seeks a set of $M$ orthogonal vectors $v$ and their associated
eigenvalues $k$ which best describe the distribution of the data.
This module takes as input the training set, and finds its
principal components (PC). The noisy test data are projected onto
the space defined by the first 20 PC, which were found to account
for about 99\% of the variance in the set, and the magnitudes of
these projections are used as attributes for the algorithm.
Experiments with larger number of PC (up to 150) showed no
significant improvement in the results.

\begin{figure}
\centering
 \includegraphics[height=65mm, width=65mm]{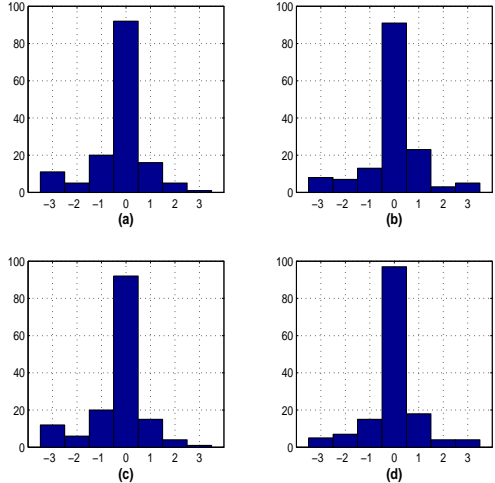}\\
 \caption{Distribution of errors in the age prediction of intermediate and old populations using the Granada models and an S/N=30.
 Error in age prediction is measured as the distance in logarithmic steps between the age of the test spectrum and the predicted age.
 Figure (a), intermediate age, and (b), old,
are the predictions of a traditional LWLR ensemble.
 Figures (c) and (d) are the predictions of our algorithm for the same ages and test spectra.} \label{f:histAgeGranRuidoNormal}
\end{figure}

\begin{figure}
\centering
 \includegraphics[height=75mm, width=85mm]{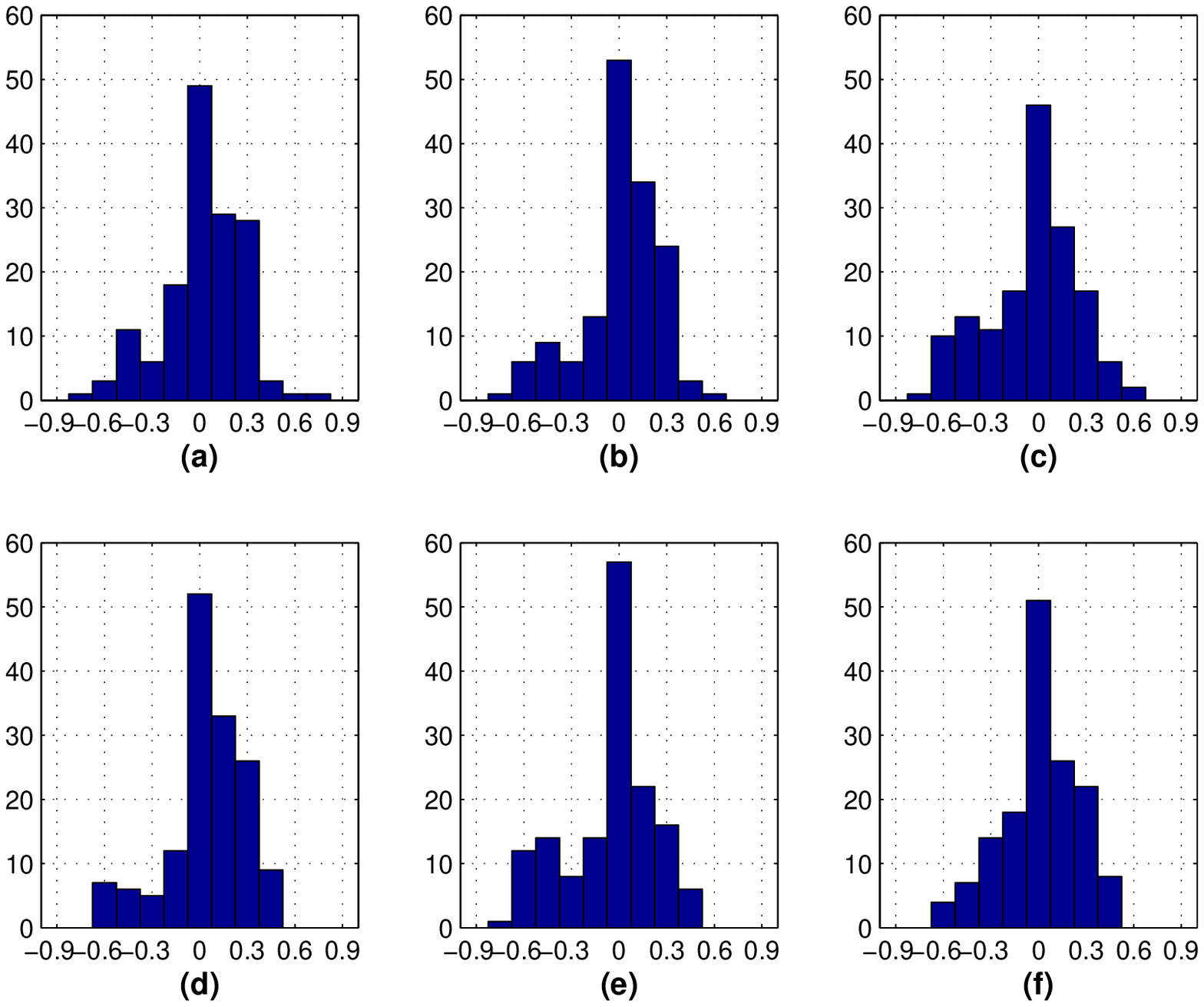}\\
 \caption{Distribution of prediction errors in the relative contribution parameters $c_{1}$, $c_{2}$ and $c_{3}$ (see
 section~\ref{s:synthetic}) using the Granada models and an S/N=30. Figures (a) to (c) are the predictions using an
 ensemble of LWLR for young, intermediate and old populations; figures (d) to (f) are predictions using our active learning
algorithm. }
  \label{f:histContGranRuidoNormal}
\end{figure}

\begin{figure}
\centering
 \includegraphics[height=75mm, width=85mm]{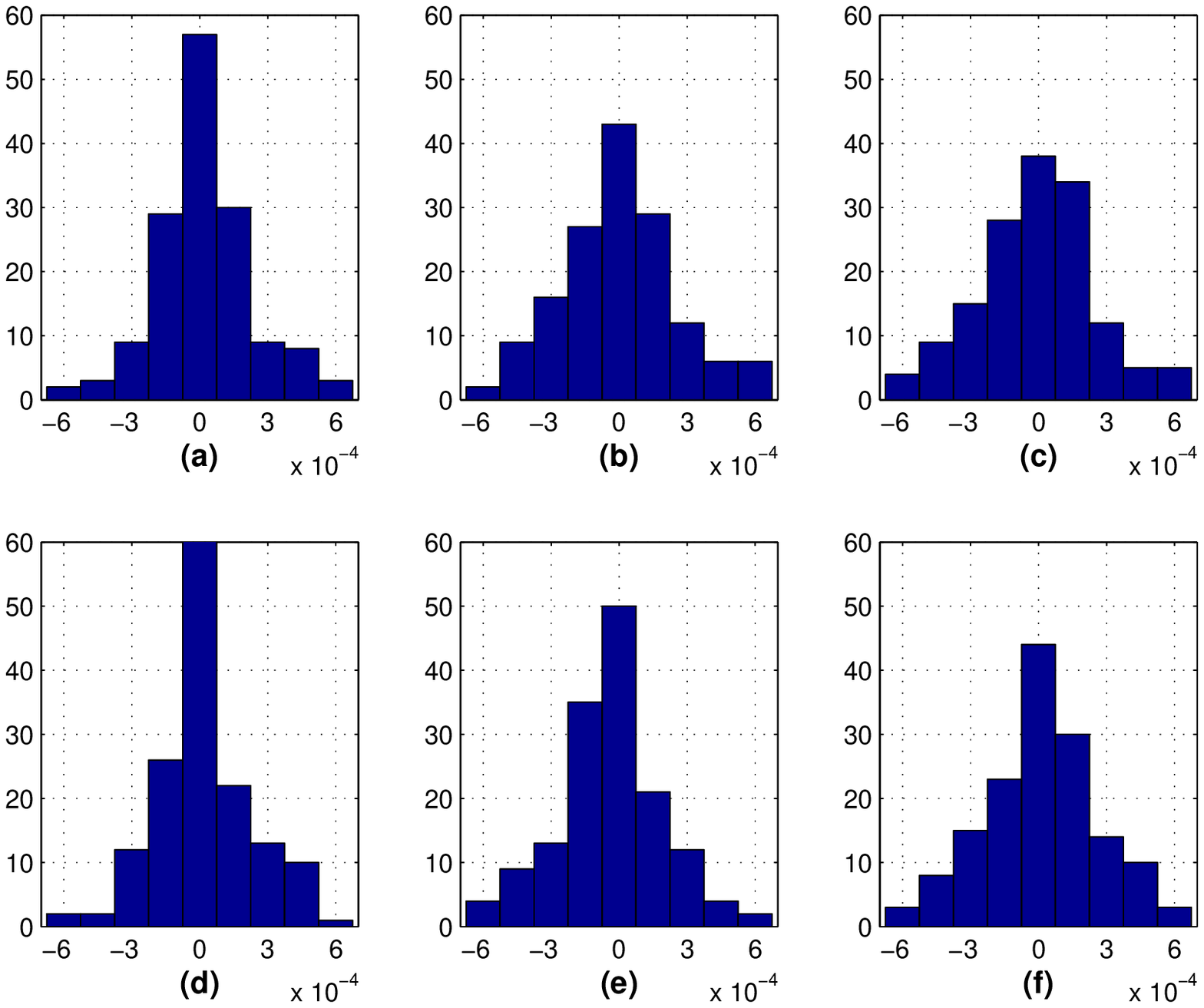}\\
 \caption{Distribution of prediction errors in the reddening parameters $r_{1}$, $r_{2}$ and $r_{3}$ (see
 section~\ref{s:synthetic}) using the Granada models and an S/N=30. Figures (a) to (c) are the predictions using an
 ensemble of LWLR for young, intermediate and old populations; figures (d) to (f) are predictions using our active learning
algorithm.}
  \label{f:histRedGranRuidoNormal}
\end{figure}

\begin{figure}
\centering
 \includegraphics[height=65mm, width=65mm]{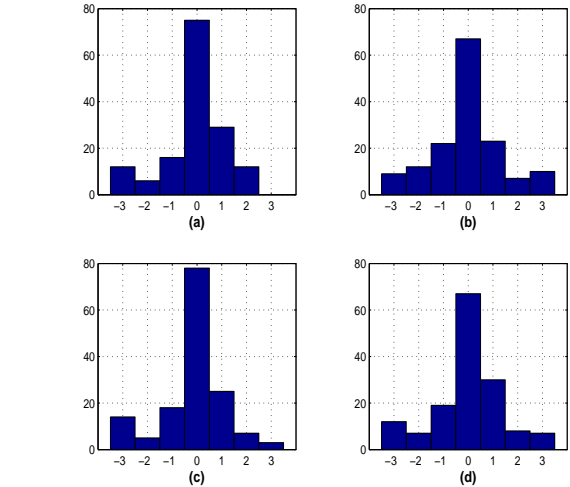}\\
 \caption{Distribution of errors in the age prediction of intermediate and old populations using the Granada models and an S/N=5.
 Figure (a), intermediate age, and (b), old, are the predictions of a traditional LWLR ensemble.
 Figures (c) and (d) are the predictions of our algorithm for the same ages and test spectra.} \label{f:histAgeGranMuchoRuido}
\end{figure}

\begin{figure}
\centering
 \includegraphics[height=75mm, width=85mm]{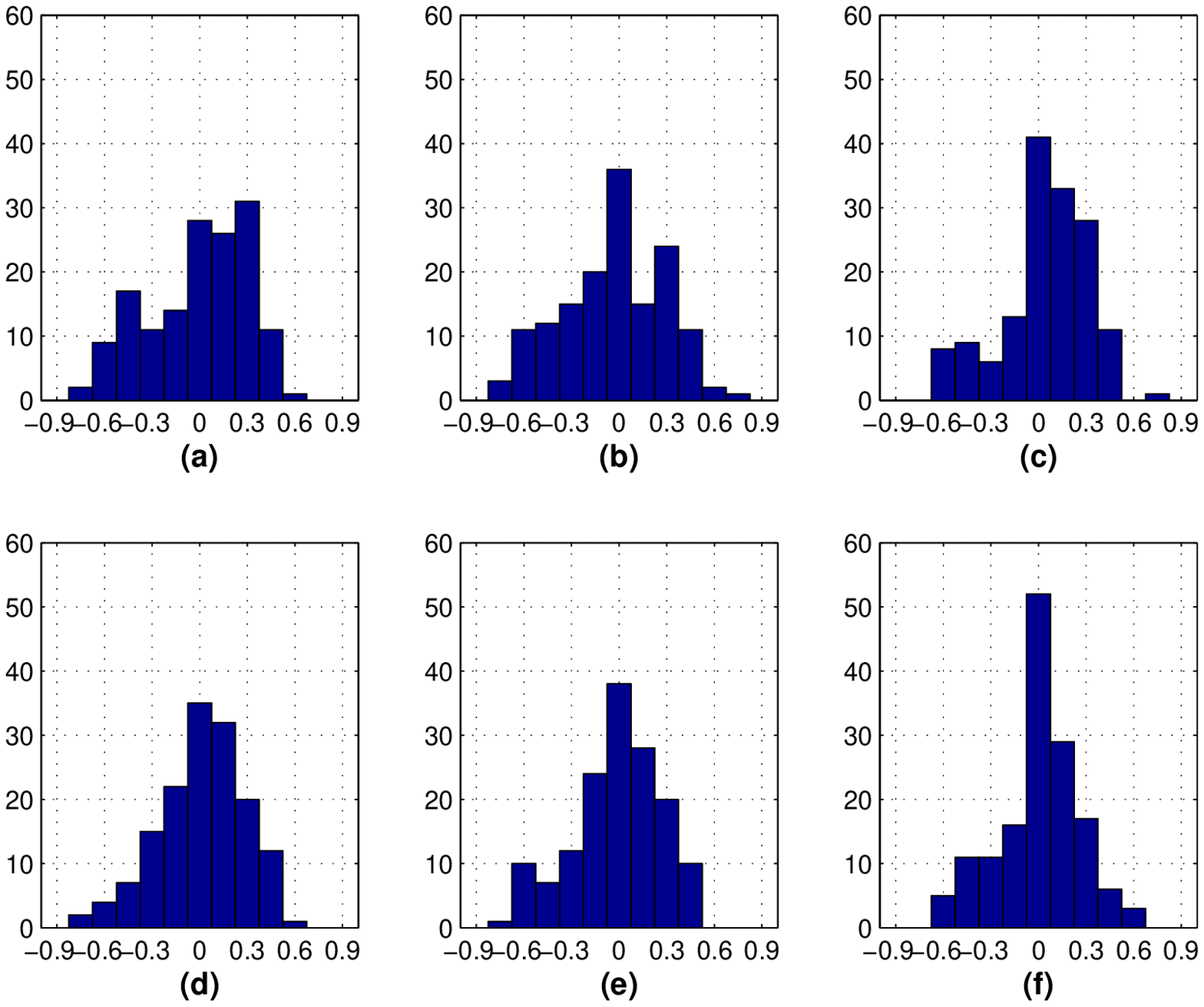}\\
 \caption{Distribution of errors in the prediction of the relative contribution parameters $c_{1}$, $c_{2}$ and $c_{3}$ (see
 section~\ref{s:synthetic}) using the Granada models and an S/N=5. Figures (a) to (c) are the predictions using an
 ensemble of LWLR for young, intermediate and old populations; figures (d) to (f) are predictions using our active learning
algorithm.}
  \label{f:histContGranMuchoRuido}
\end{figure}

\begin{figure}
\centering
 \includegraphics[height=75mm, width=85mm]{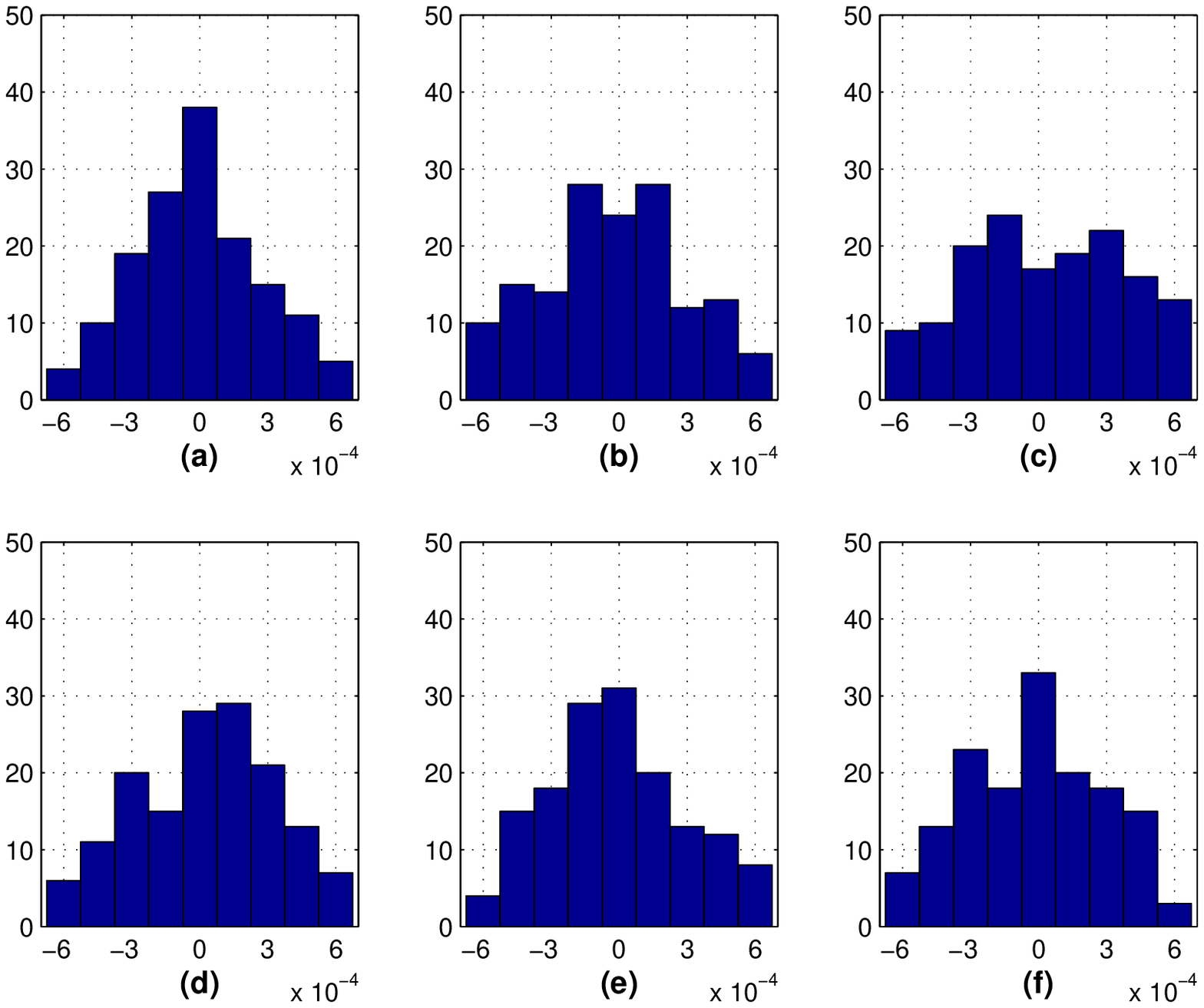}\\
 \caption{Distribution of prediction errors in the reddening parameters $r_{1}$, $r_{2}$ and $r_{3}$ (see
 section~\ref{s:synthetic}) using the Granada models and an S/N=5. Figures (a) to (c) are the predictions using an
 ensemble of LWLR for young, intermediate and old populations; figures (d) to (f) are predictions using our active learning
algorithm. }
  \label{f:histRedGranMuchoRuido}
\end{figure}

\begin{figure}
\centering
 \includegraphics[height=105mm, width=85mm]{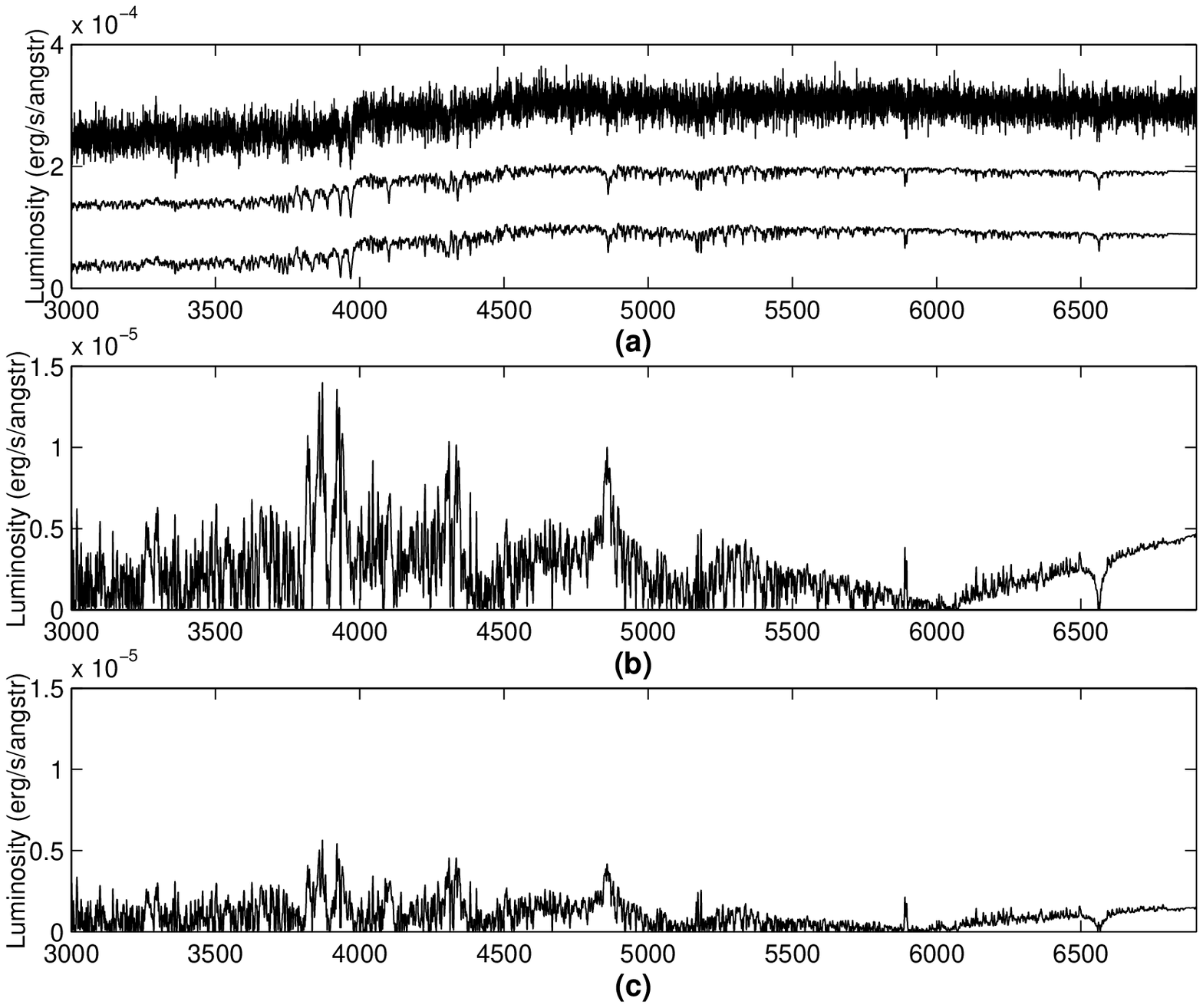}\\
 \caption{Graphical comparison of results using the Granada models and noisy data, ration S/N=5. Figure (a) from top to bottom
 and shifted by a constant to aid visualization: noisy test spectrum,
 spectrum recovered using ensemble of LWLR and spectrum recovered using active learning. Figure (b) show the residuals of the
 reconstructed spectrum using ensemble of LWLR and figure (c) is the corresponding residuals of using the active learning technique.}
  \label{f:granMuchoRuido}
\end{figure}

\begin{figure}
\centering
 \includegraphics[height=105mm, width=85mm]{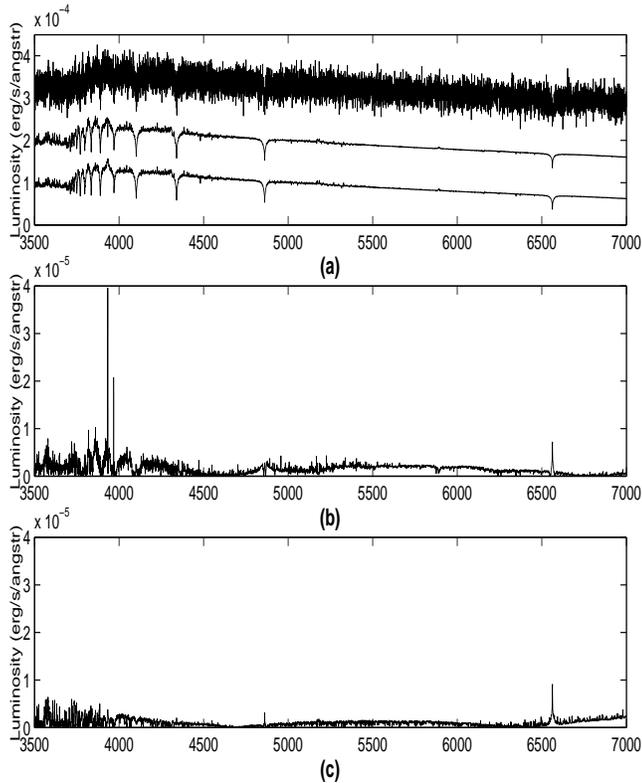}\\
 \caption{Same as Figure~\ref{f:granMuchoRuido} but using the Padova models and noisy data, ratio S/N=5. Figure (a) from top to bottom
 and shifted by a constant to aid visualization: noisy test spectrum,
 spectrum recovered using ensemble of LWLR and spectrum recovered using active learning. Figure (b) show the residuals of the
 reconstructed spectrum using ensemble of LWLR and figure (c) is the corresponding residuals of using the active learning technique.}
  \label{f:padMuchoRuido}
\end{figure}

Figure~\ref{f:histAgeGranRuidoNormal} shows the error distribution
in the age prediction using noisy data ($S/N=30$) with an ensemble
of LWLR and active learning using the Granada models. In the case
of intermediate age prediction, both algorithms achieve almost
identical errors. In contrast, for prediction of old populations
the active learning algorithm slightly outperforms the ensemble of
LWLR. It is important to note that the central peak contains more
than 60\% of the results, while the +1, -1 bins include about 20\%
of the cases. For our method, about 85\% shows an error in the age
determination that is equal or smaller than one age step.
Prediction of the relative contributions, presented in
Figure~\ref{f:histContGranRuidoNormal}, is not as peaked as the
age prediction but still a substantial fraction is inside a small
error. Our method in this case shows a moderate improvement with
respect to the LWLR. This same behavior can be observed in
Figure~\ref{f:histRedGranRuidoNormal} where error distribution in
the prediction of reddening parameters $r_{1}$,$r_{2}$ and $r_{3}$
are presented. Results for the Padova theoretical models are
similar to those for the Granada models, although the improvement
achieved by our active learning technique is much higher in the
case of the Padova models, specially for the
parameters of the old populations. 
Another set of figures presents results of experiments with very
noisy data, using an S/N=5. For the Granada models distribution of
errors in predictions are shown in
Figures~\ref{f:histAgeGranMuchoRuido},
\ref{f:histContGranMuchoRuido} and \ref{f:histRedGranMuchoRuido}.
It is remarkable that even with low quality (S/N=5) data the
algorithm does such a good estimate of the population ages. For
our method in about 80\% of the cases the error in the age
determination is equal or less than one age step. However, it is
evident that the active technique was unable to improve accuracy
due to high levels of noise in some particular cases. For
instance, for the Granada models prediction of reddening parameter
for young populations, $r_{1}$, presented higher error rates using
our algorithm than using a traditional LWLR ensemble. However, in
the estimation of reddening for intermediate and old populations
the inverse of this occurred, the traditional approach was
outperformed by our algorithm. Our algorithm also achieved higher
accuracy for the estimation of relative contribution parameters.
For the Padova models, in the majority of the cases better results
were achieved by the active algorithm; only in one case, the age
prediction of old
populations, the active algorithm had slightly higher errors. 
In Figures \ref{f:granMuchoRuido} and \ref{f:padMuchoRuido} we
show graphical comparisons between test spectra and reconstructed
ones using an LWLR ensemble and our active learning algorithm with
an S/N=5. Comparing these figures with the results obtained for
noiseless data we can say that the improvement in the fitting of
the active technique is lower with very noisy data, although the
advantage of the technique is still significant at the lowest S/N
ratio.

\subsection{Lick Indices}
\begin{center}
\begin{figure}
 \includegraphics[height=85mm, width=85mm]{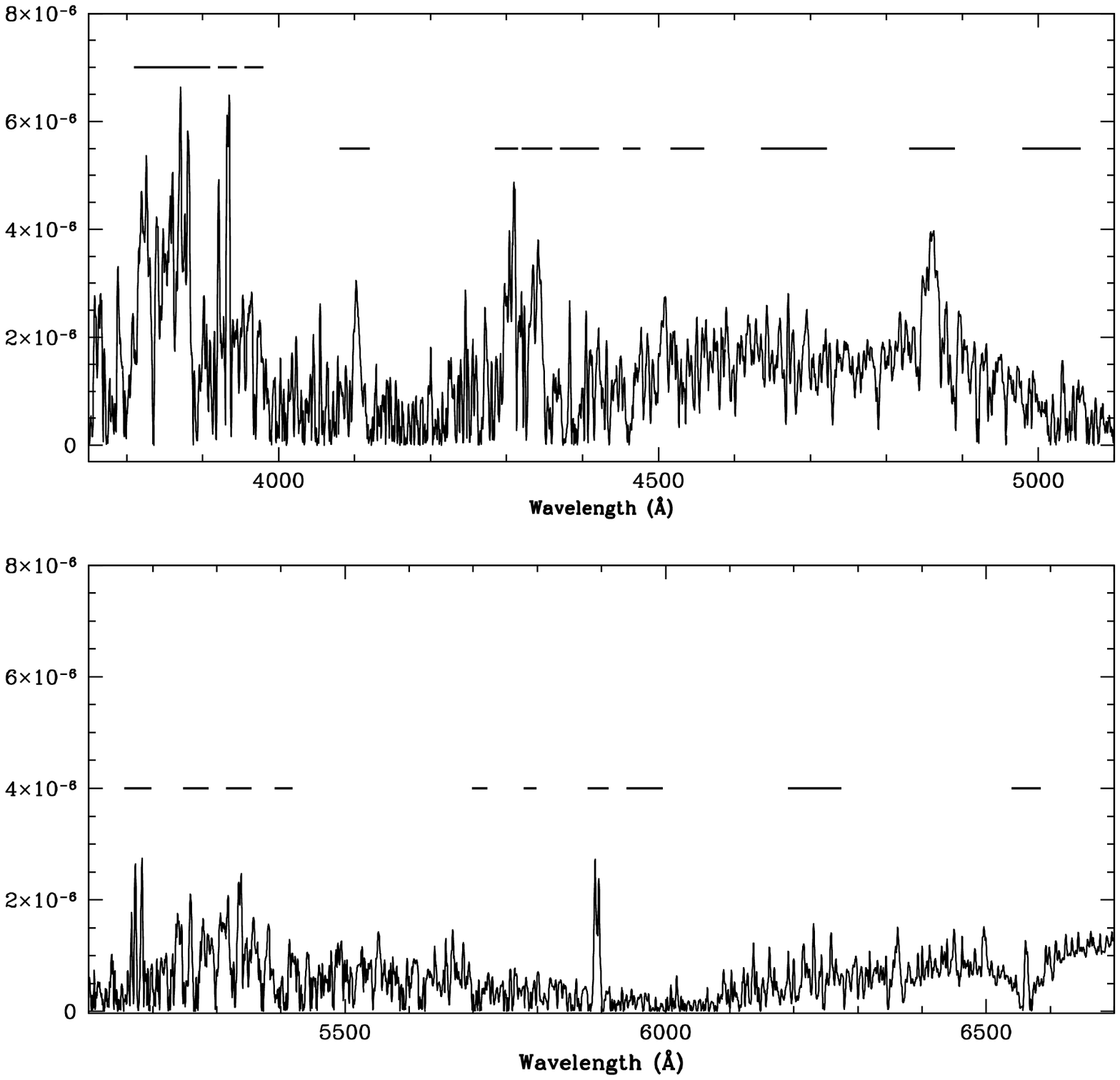}\\
  \caption{Residuals between a test and a predicted spectrum, the horizontal bands above the
  residuals show the Lick Indices. We split the spectrum to aid visualization, the top of the figure shows residuals in the blue part while
  the bottom shows the red part residuals. }
  \label{f:Lick}
\end{figure}
\end{center}

One remarkable aspect found in the experiments with noise included
is that even when using a large number of PC the residuals showed
relatively high peaks in some specific narrow spectral regions.
Surprisingly several of these high residual regions coincide with
the central band of the Lick indices (listed in
Table~\ref{t:Lick}), see Figure~\ref{f:Lick} where the Lick
indices are superimposed on the residuals of the reconstructed
spectrum. In other words, giving equal weights to all pixels (or
fluxes) produced larger residuals located in these narrow regions.

\begin{figure}
\centering
 \includegraphics[height=65mm, width=65mm]{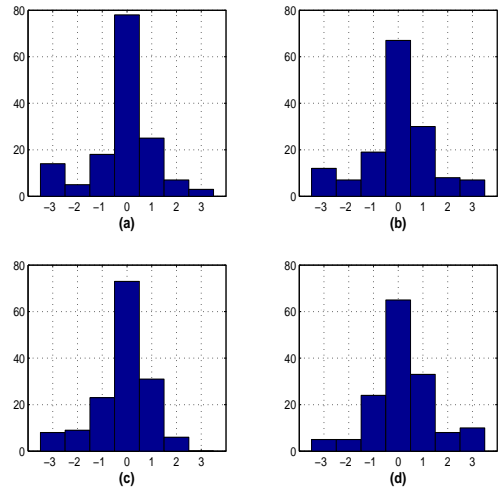}\\
 \caption{Distribution of errors in the age prediction of intermediate and
 old populations for the Granada models using an S/N=5. Figure (a), intermediate age, and (b), old,
 are the predictions of active learning and the original data.  Figures (c) and (d) are the predictions of
 active learning using only the fluxes of the central lines of the Lick indices. } \label{f:histAgeLinGranMuchoRuido}
\end{figure}

\begin{figure}
\centering
 \includegraphics[height=75mm, width=85mm]{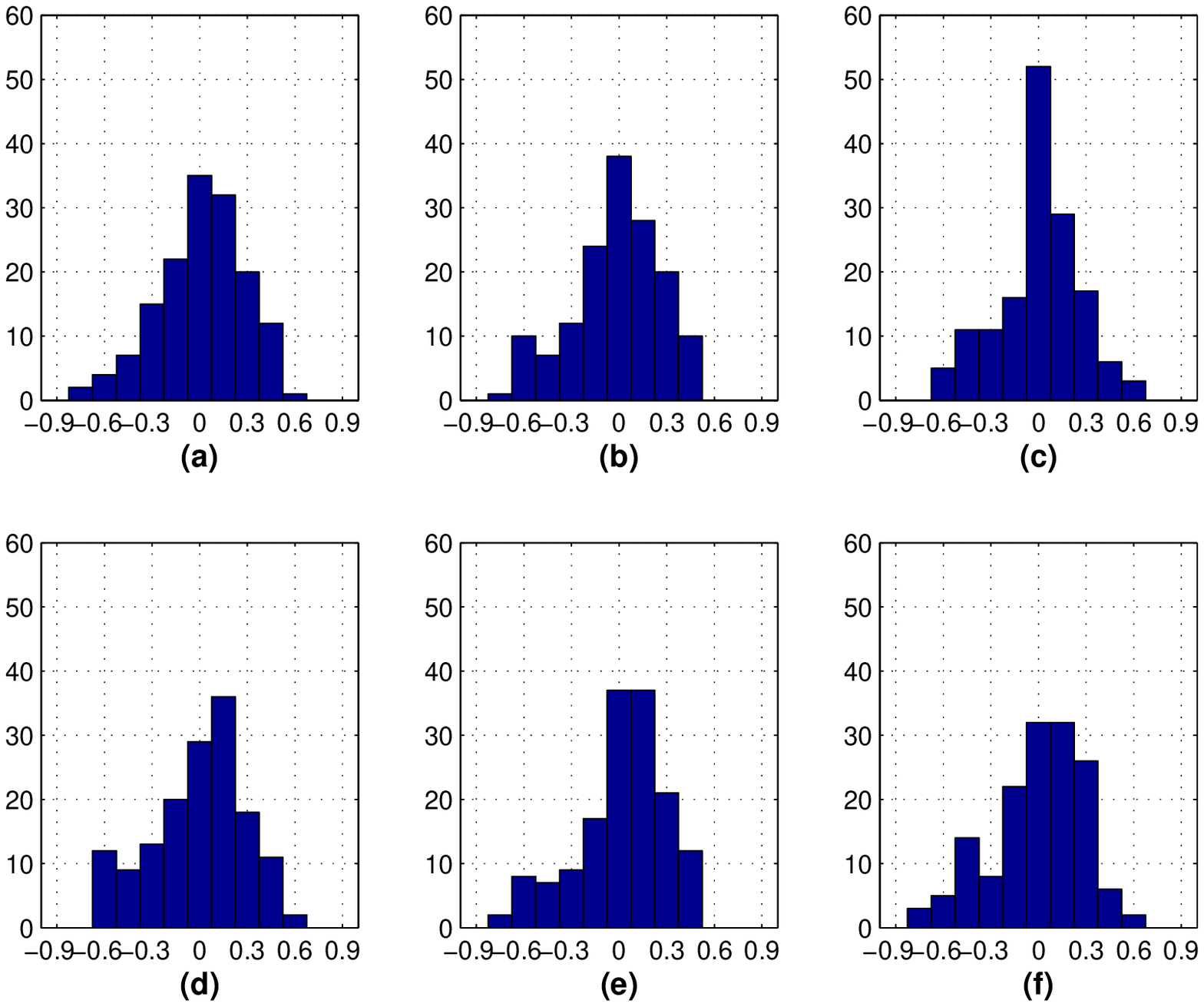}\\
 \caption{Distribution of prediction errors in the relative contribution parameters $c_{1}$, $c_{2}$ and $c_{3}$
 (see section \ref{s:synthetic}) for the Granada models using an S/N=5. Figures (a) to (c) are the predictions
 of active learning and the original data for young, intermediate and old populations respectively. Figures
 (d) and (f) are the predictions of active learning using only the fluxes of the central lines of the Lick indices. }
  \label{f:histContLinGranMuchoRuido}
\end{figure}

\begin{figure}
\centering
 \includegraphics[height=75mm, width=85mm]{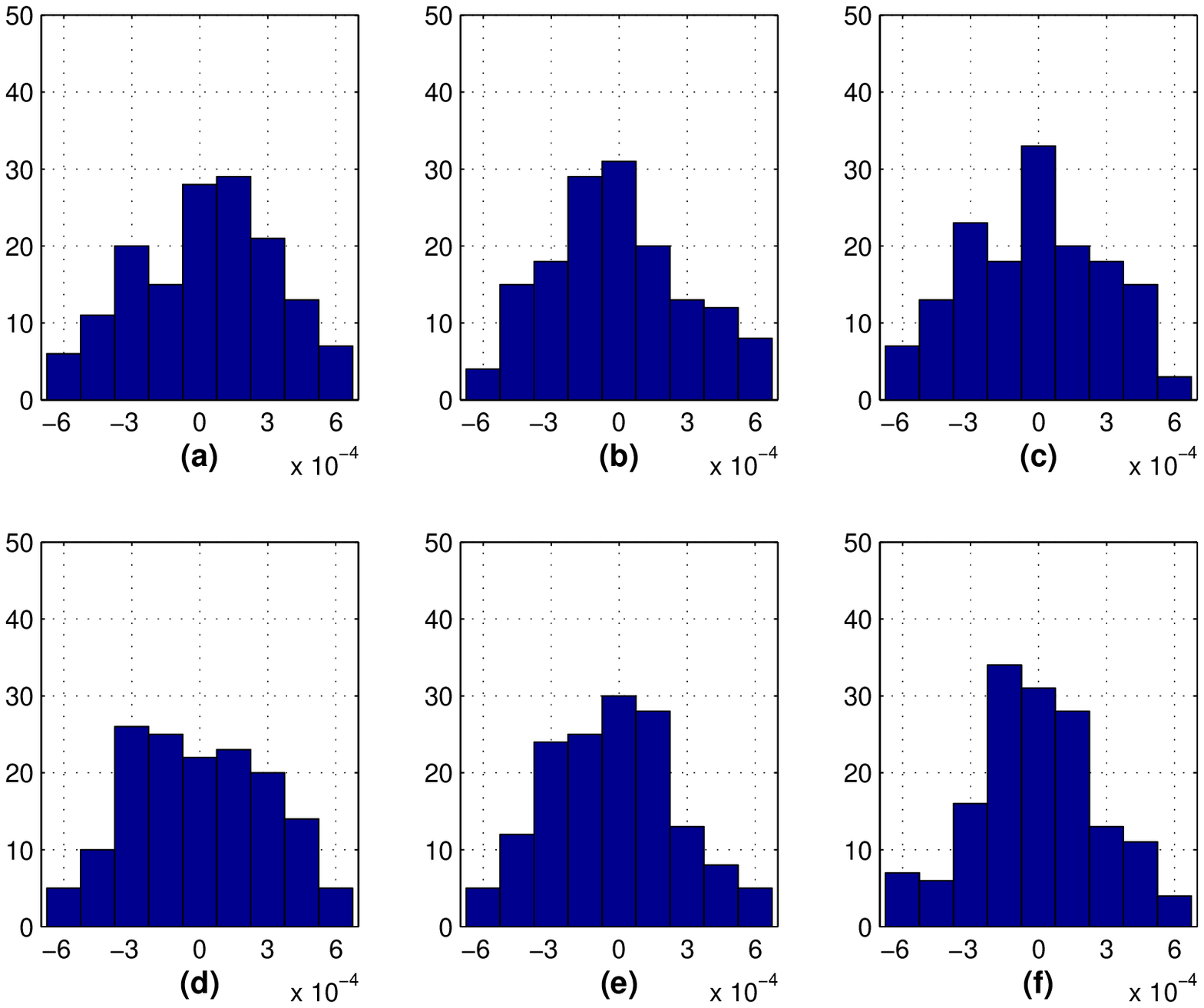}\\
 \caption{Distribution of prediction errors in the reddening parameters $r_{1}$, $r_{2}$ and $r_{3}$ (see section \ref{s:synthetic})
 for the Granada models using an S/N=5. Figures (a) to (c) are the predictions of active learning and the original
 data for young, intermediate and old populations respectively.  Figures (d) and (f) are the predictions of active
 learning using only the fluxes of the central lines of the Lick indices.}
  \label{f:histRedLinGranMuchoRuido}
\end{figure}

We opted to explore whether prior knowledge about the Lick indices
can help machine learning algorithms to provide a more accurate
prediction. Thus, we experimented using two different approaches
aimed at giving more influence to the central bands of the Lick
indices. In the first approach we discarded all information about
most of the spectra, keeping only the flux information
corresponding to the central bands of the Lick indices. The
learning algorithm thus predicts the reddening parameters using
only this reduced subset of fluxes. In a similar way, the
contribution of ages is estimated using the same subset of fluxes.
Figures \ref{f:histAgeLinGranMuchoRuido},
\ref{f:histContLinGranMuchoRuido} and
\ref{f:histRedLinGranMuchoRuido} show a comparison of error
distributions between active learning when using the original data
and active learning when using the Lick indices for the Granada
models. We present here only the results using very noisy data
(S/N =5) given that previous experiments showed higher error rates
for this scenario. For the Padova models using this prior
knowledge did not yield higher accuracy in the case of an LWLR
ensemble; predictions from active learning using the original data
are more accurate. However, when using the Lick indices and active
learning reddening parameters are estimated better, as well as the
relative contribution of ages. In the case of the Granada models
the best results were achieved by active learning using the
original data.

\begin{table}
\centerline{
  \begin{tabular}{lrr}
   \hline
 Name     &   Index Begin & Index End\\
 \hline
B\&H\_CNB   &  3810.0  &    3910.0\\
HKratioK  &  3920.0   &    3945.0\\
 HKratioH  &  3955.0  &     3980.0\\
 Hd & 4080.0 & 4120.0\\
  Lick\_CN1 & 4143.375 &   4178.375\\
   B\&H\_CaI &    4215.0 & 4245.0\\
   Lick\_Ca4227 & 4223.500 & 4236.000\\
    Lick\_G4300 & 4282.625 & 4317.625 \\
    B\&H\_G & 4285.0 & 4315.0\\
     Hg &     4320.0  &    4360.0\\
     Lick\_Fe4383 & 4370.375 & 4421.625\\
      Lick\_Ca4455 & 4453.375 &   4475.875\\
Lick\_Fe4531 & 4515.500 & 4560.500\\
Lick\_C4668 & 4635.250 & 4721.500\\
 B\&H\_Hb &     4830.0 & 4890.0\\
Lick\_Hb & 4848.875 &   4877.625\\
 Lick\_Fe5015 & 4979.000 & 5055.250\\
Lick\_Mg2 & 5155.375  &  5197.875\\
 Lick\_Fe527 & 5247.375 & 5287.375\\
 Lick\_Fe5335 & 5314.125  &  5354.125\\
  Lick\_Fe5406 & 5390.250 & 5417.750\\
Lick\_Fe5709 & 5698.375 &   5722.125\\
 Lick\_Fe5782 & 5778.375 & 5798.375\\
  Lick\_NaD & 5878.625 &   5911.125\\
   Lick\_TiO1 & 5938.875 & 5995.875\\
Lick\_TiO2 & 6191.375 &   6273.875\\
 Ha & 6540.0 & 6585.0\\
\hline
\end{tabular}}
\caption{ The table was constructed based in the SLOANE index
list. We remove the duplicate indices leaving where possible the
Lick indices. We also added indices for the Balmer lines, the \Ha,
\Hg~ and \Hd. The HK ratio index was decomposed into two bands. By
Index here we mean only the central band and not the continuum
side bands.}
 \label{t:Lick}
\end{table}

\begin{figure}
\centering
 \includegraphics[height=65mm, width=65mm]{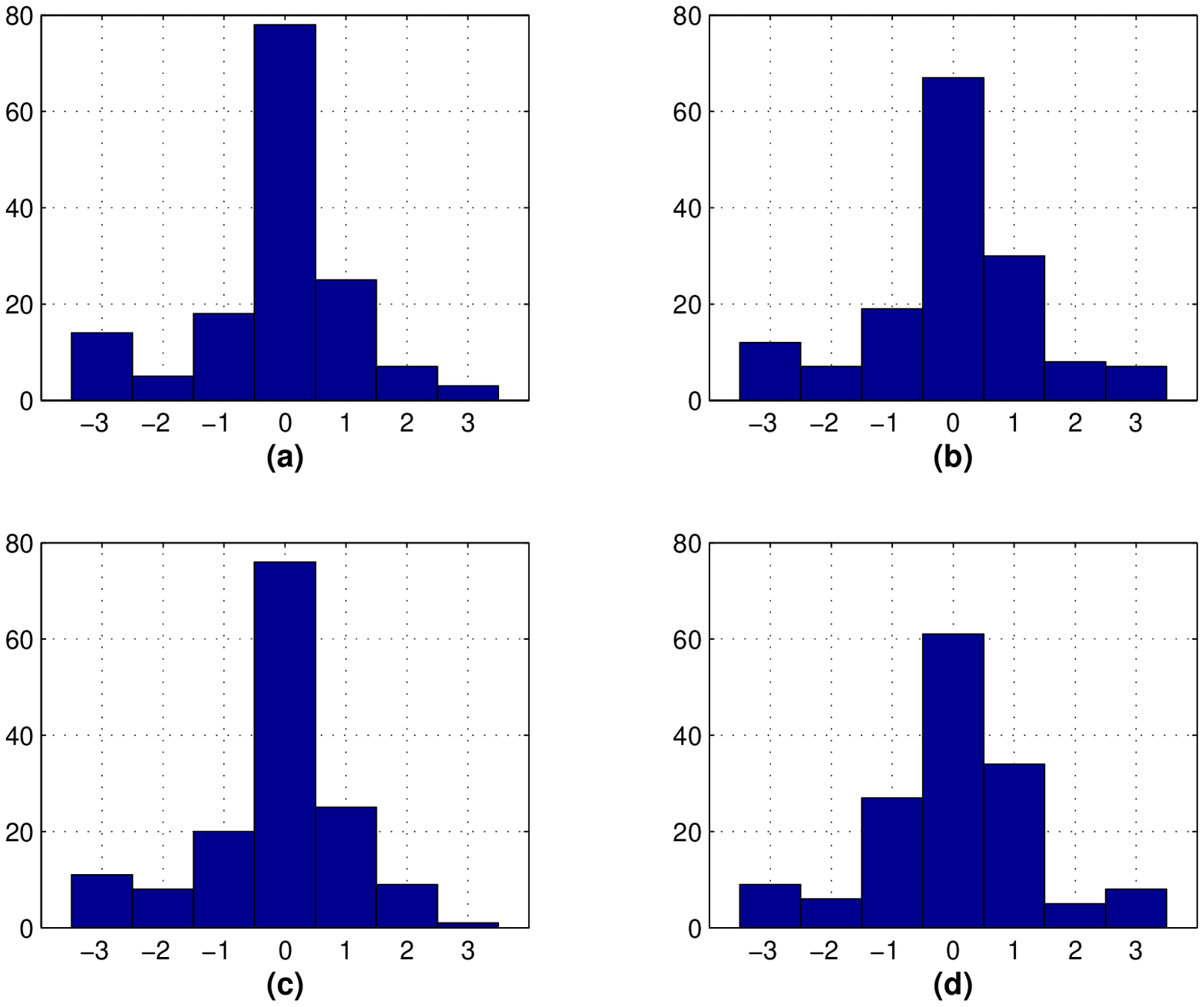}\\
 \caption{Distribution of prediction errors in the age prediction of intermediate and
 old populations for the Granada models using an S/N=5. Figure (a), intermediate age, and (b), old,
 are the predictions of active learning and the original data.  Figures (c) and (d) are the predictions of
 active learning with the fluxes of the central lines of the Lick indices magnified by a constant $k=4$. } \label{f:histAgeKGranMuchoRuido}
\end{figure}

\begin{figure}
\centering
 \includegraphics[height=75mm, width=85mm]{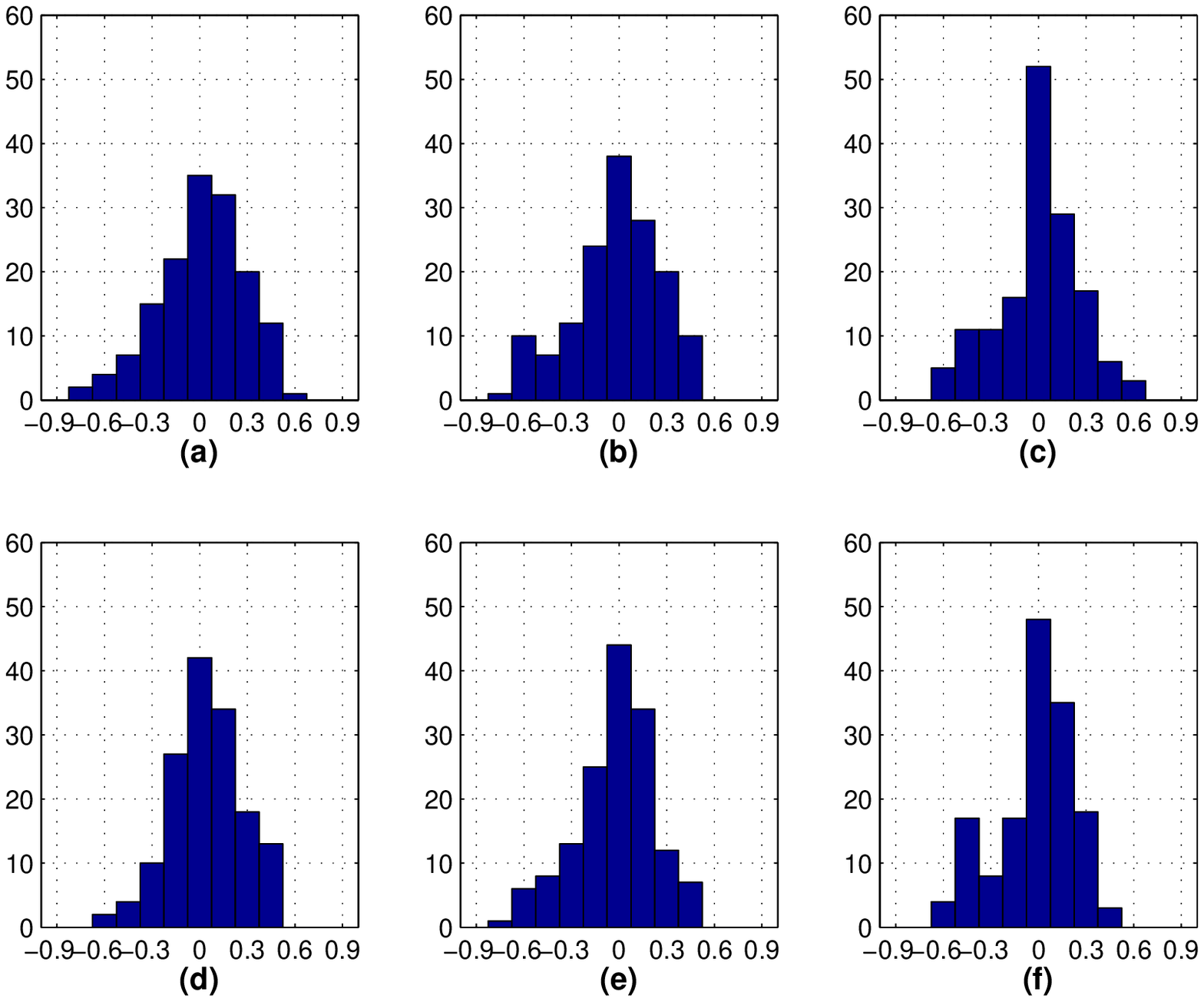}\\
 \caption{Distribution of prediction errors in the relative contribution parameters $c_{1}$, $c_{2}$ and $c_{3}$ (see section \ref{s:synthetic})
 for the Granada models using an S/N=5. Figures (a) to (c) are the predictions of active learning and the original data
 for young, intermediate and old populations respectively.  Figures (d) and (f) are the predictions of active learning
 with the fluxes of the central lines of the Lick indices magnified by a constant $k=4$.} \label{f:histContKGranMuchoRuido}
\end{figure}

\begin{figure}
\centering
 \includegraphics[height=75mm, width=85mm]{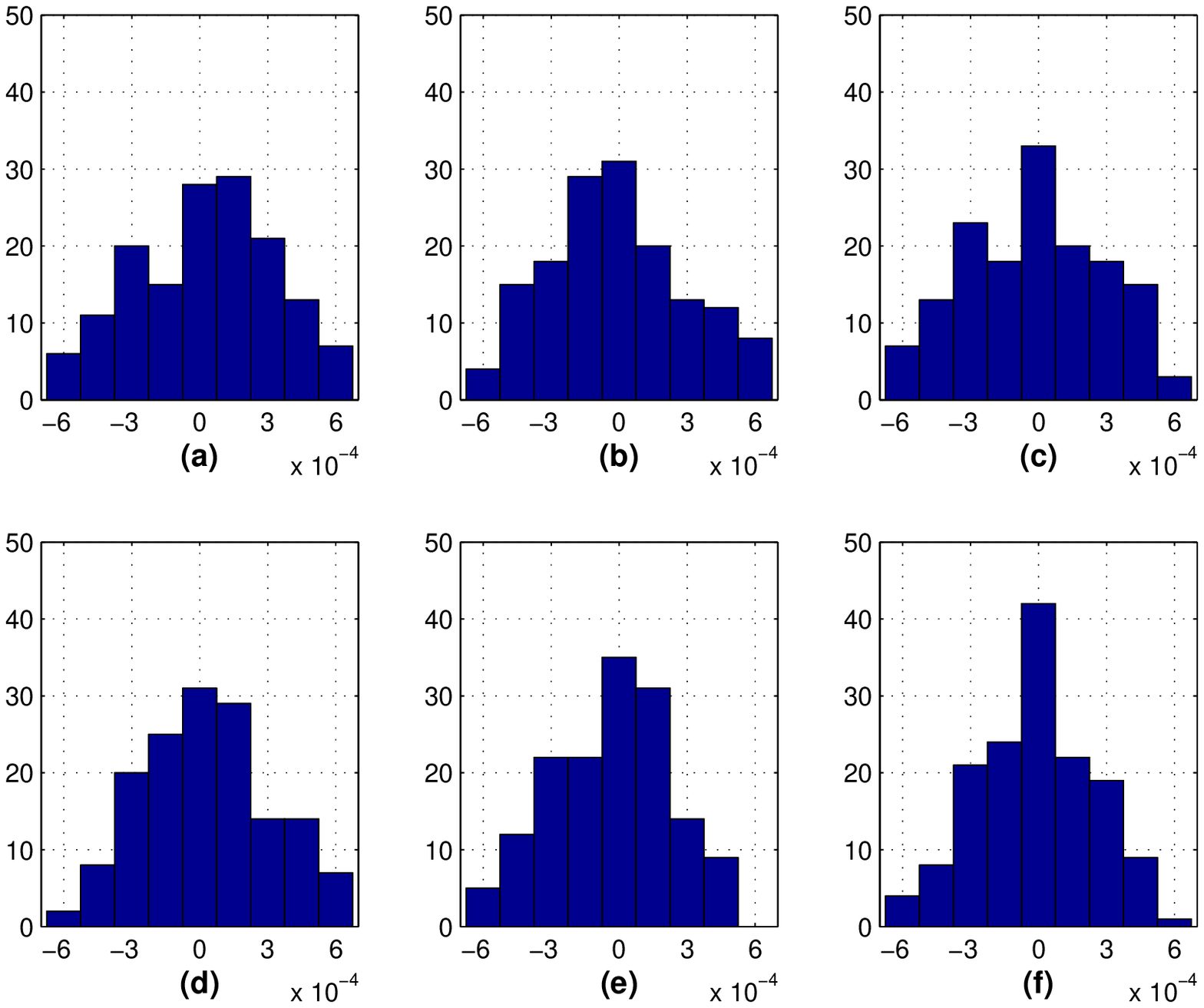}\\
 \caption{Distribution of prediction errors in the reddening parameters $r_{1}$, $r_{2}$ and $r_{3}$ (see section \ref{s:synthetic})
 for the Granada models using an S/N=5. Figures (a) to (c) are the predictions of active learning and the original data
 for young, intermediate and old populations respectively.  Figures (d) and (f) are the predictions of
 active learning with the fluxes of the central lines of the Lick indices magnified by a constant $k=4$.} \label{f:histRedKGranMuchoRuido}
\end{figure}

The other approach for incorporating prior knowledge consists of
increasing the relevance of the Lick indices. By doing so,
differences in the Lick indices of the data will have more weight
than the differences through the rest of the spectrum; this will be
reflected when LWLR selects the closest examples to the test
spectrum (see Subsection~\ref{s:LWLR}). To do this we multiplied
the energy fluxes in the wavelengths corresponding to the Lick
indices by a constant $k$. That is, fluxes in regions defined by
Lick indices where deemed to be $k$ times more important than
pixels in other regions. This value of $k=4$ was set
experimentally with a 10-fold cross-validation procedure. We
present results using very noisy data (S/N=5). These results are
similar to results previously discussed. We find that while for
the Padova models the best results come from active learning with
prior knowledge, for the Granada models this is not the case and
the best results come from active learning and the original data.
Figures \ref{f:histAgeKGranMuchoRuido} to
\ref{f:histRedKGranMuchoRuido} present error distribution of these
experiments. This whole topic will be further investigated in a
forthcoming paper, where the method is applied to real data.

Using prior knowledge did not yield meaningful improvements,
moreover, for some parameters the error increased when
incorporating prior knowledge. In another attempt to improve
results with very noisy data we carried out another set of
experiments. This time we build an ensemble using the predictions
from the three approaches: active learning using the original
data, active learning using the fluxes corresponding to the Lick
indices and active learning with more weight given to fluxes
around the central bands of the Lick indices. All the ensemble
predictions are then computed as the average of the predictions
from each approach.

\begin{figure}
\centering
 \includegraphics[height=65mm, width=65mm]{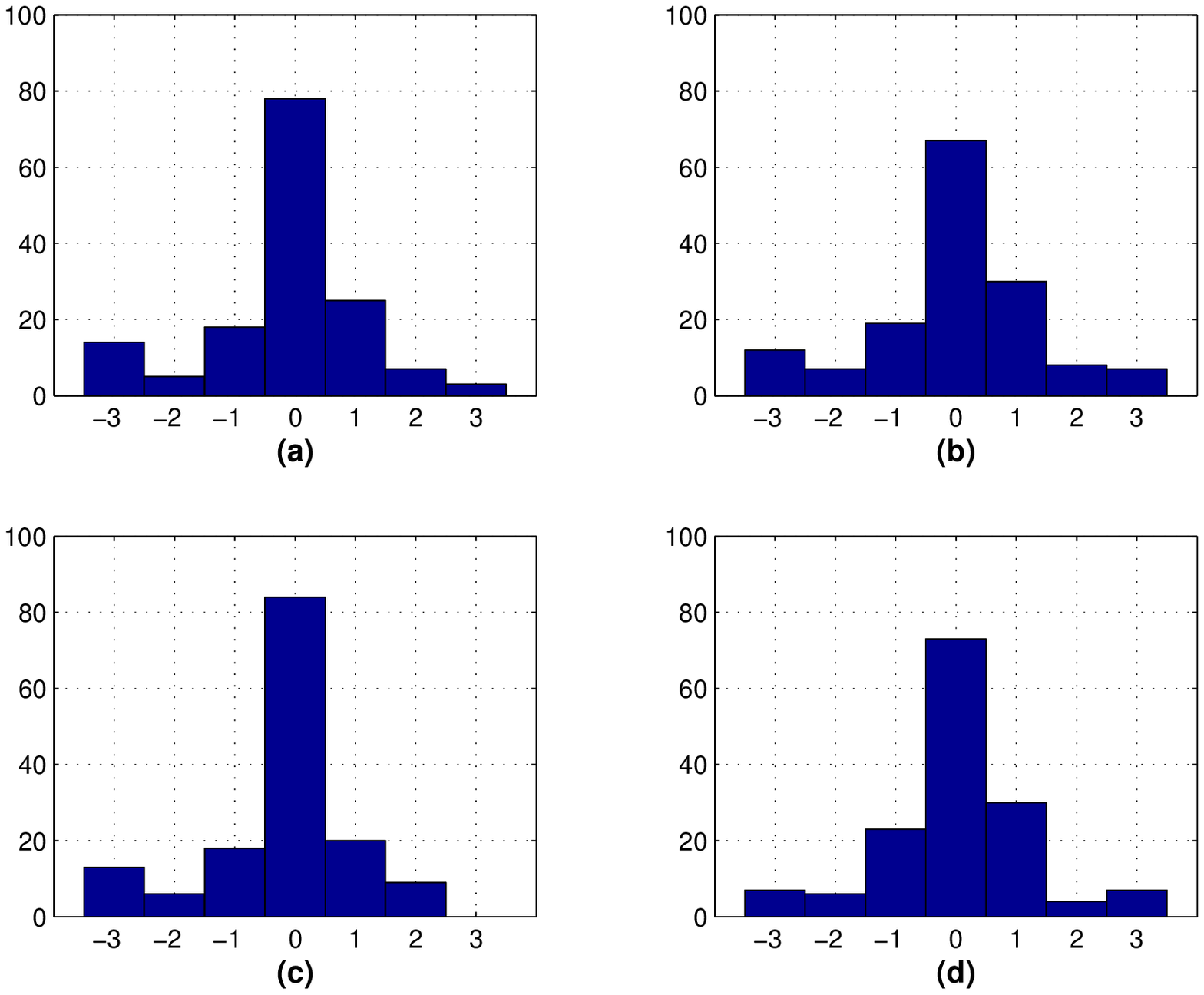}\\
 \caption{Distribution of prediction errors in the age prediction of intermediate and old populations for the Granada
 models using an S/N=5. Figure (a), intermediate age, and (b), old, are the predictions of
the active learning algorithm. Figures (c) and (d) are the
predictions of the ensemble combining prior knowledge. }
\label{f:histAgeEnsGranMuchoRuido}
\end{figure}

\begin{figure}
\centering
 \includegraphics[height=75mm, width=85mm]{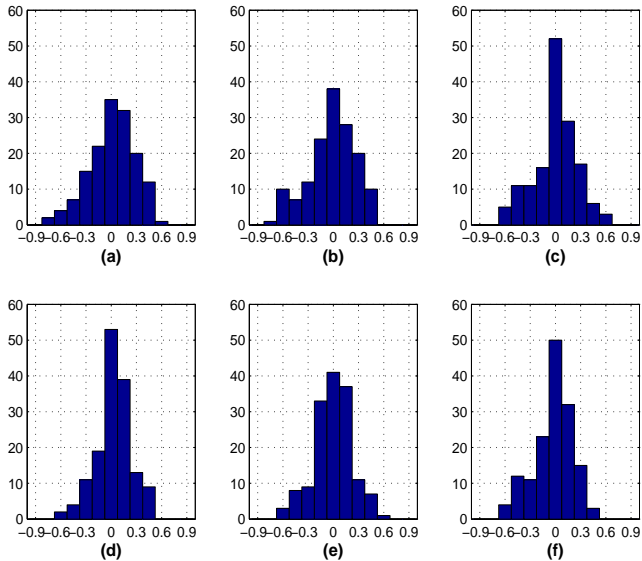}\\
 \caption{Distribution of prediction errors in the relative contribution parameters $c_{1}$, $c_{2}$ and $c_{3}$ (see
 section~\ref{s:synthetic}) for the Granada models using an S/N=5. Figures (a) to (c) are the predictions of active
 learning and the original data for young, intermediate and old populations respectively. Figures (d) and (f) are
 the predictions of the active learning ensemble combining prior knowledge.}
  \label{f:histContEnsGranMuchoRuido}
\end{figure}

\begin{figure}
\centering
 \includegraphics[height=75mm, width=85mm]{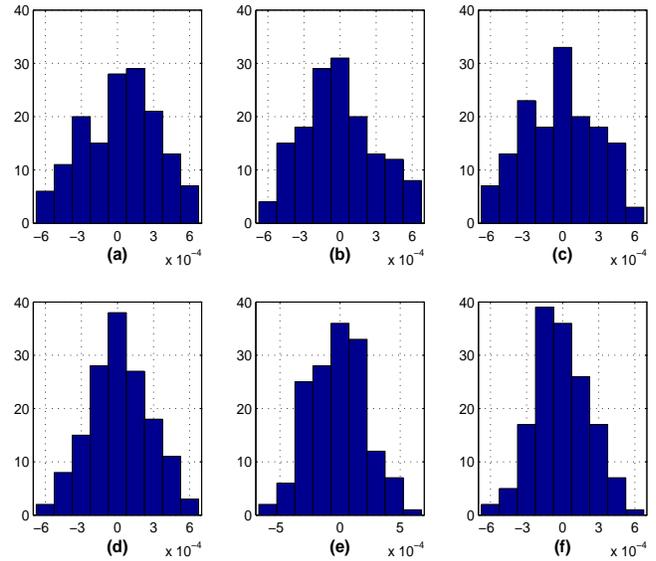}\\
 \caption{Distribution of prediction errors in the reddening parameters $r_{1}$, $r_{2}$ and $r_{3}$ (see section \ref{s:synthetic})
 for the Granada models using an S/N=5. Figures (a) to (c) are the predictions of active
 learning and the original data for young, intermediate and old populations respectively, figures (d) to (f) are predictions
 of the ensemble combining prior knowledge. }
  \label{f:histRedEnsGranMuchoRuido}
\end{figure}

\begin{figure}
\centering
 \includegraphics[height=105mm, width=85mm]{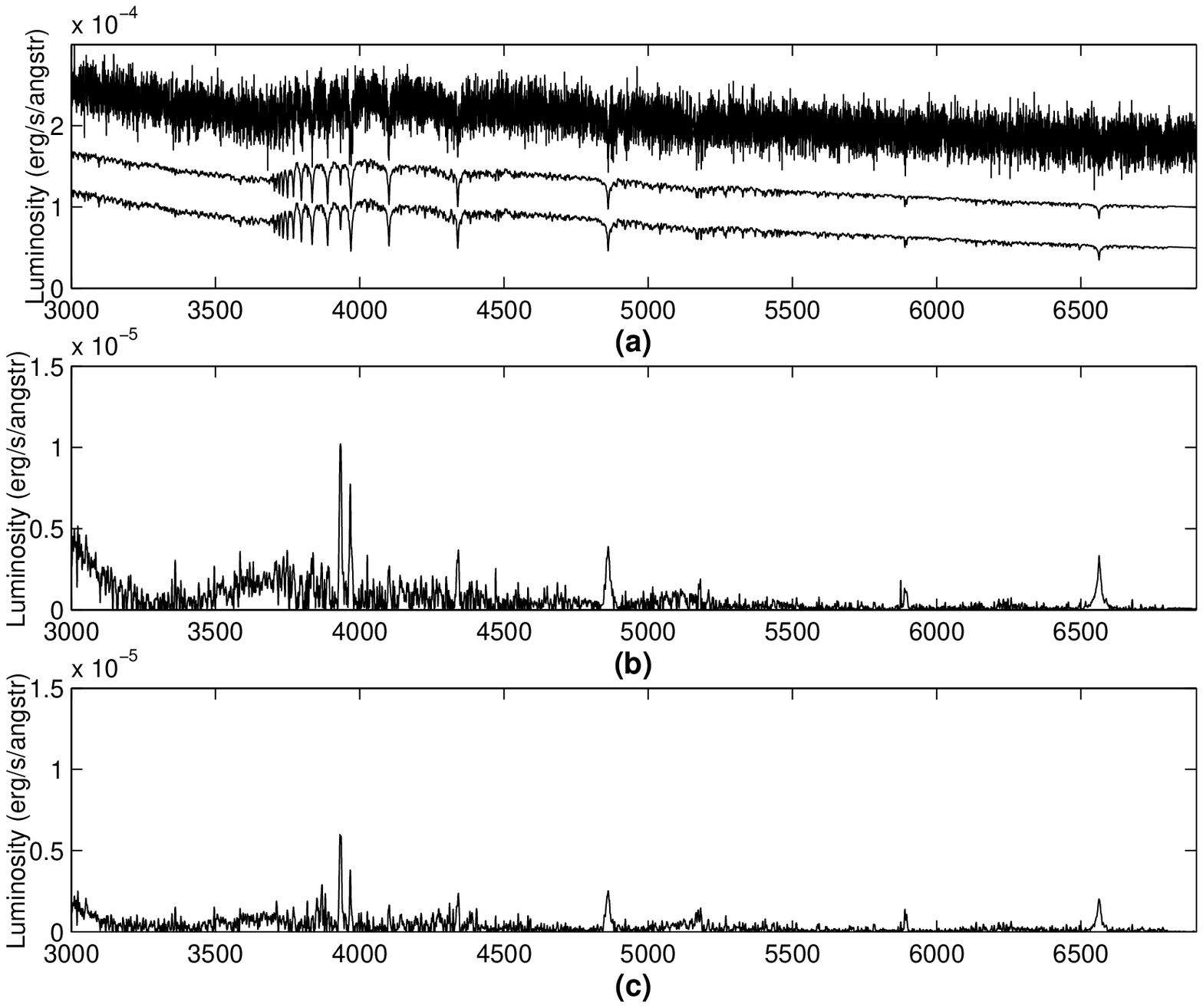}\\
 \caption{Graphical comparison of results using the Granada models and noisy data, ratio S/N=5.
Figure (a) from top to bottom and shifted by a constant to aid
visualization: noisy test spectrum, spectrum recovered using
active learning and the original data and spectrum recovered using
active learning combining predictions. Figures (b) and (c) show
the relative difference between test spectrum and predicted
spectra in the same listed order.}
  \label{f:ensGran}
\end{figure}

\begin{figure}
\centering
 \includegraphics[height=105mm, width=85mm]{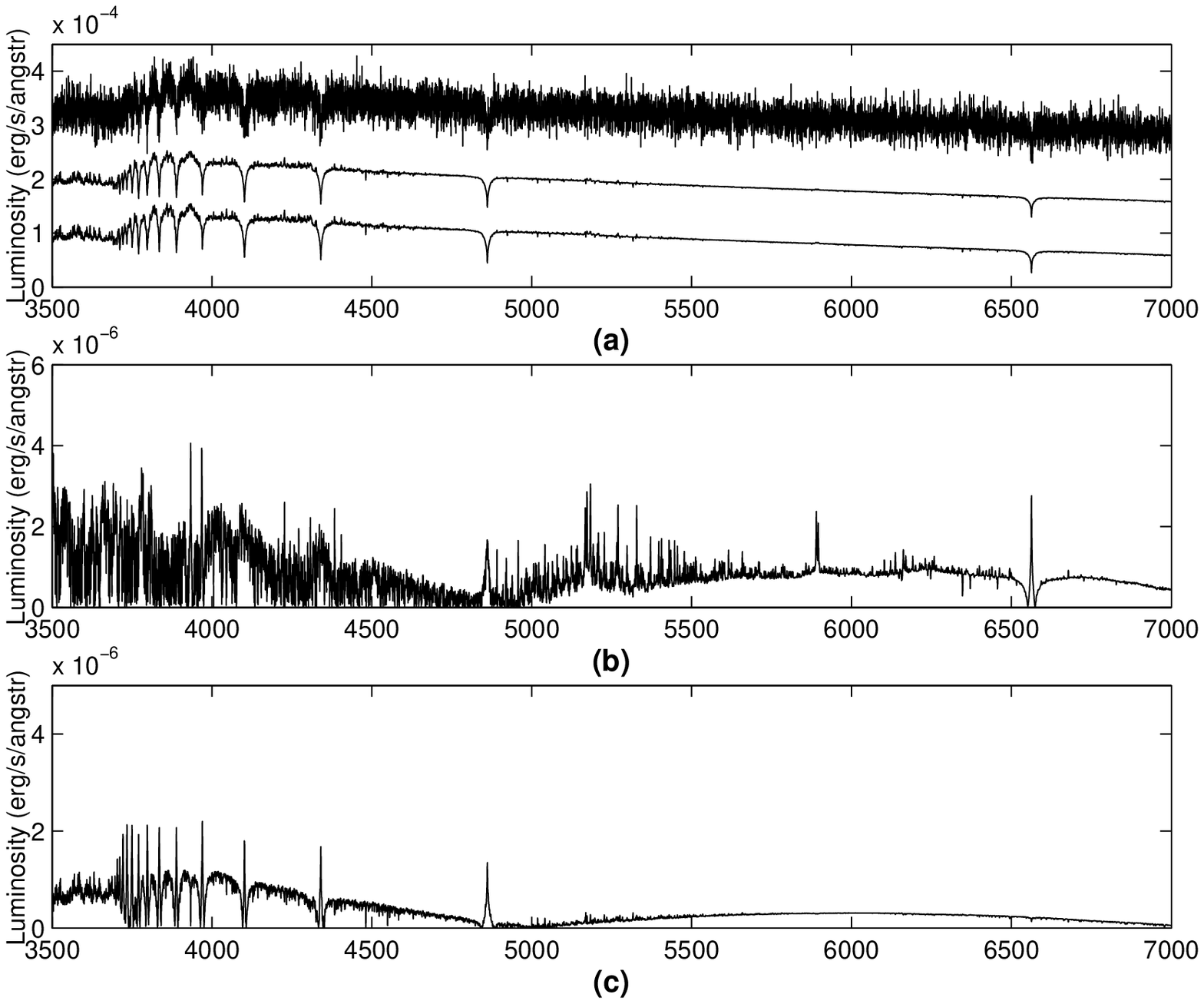}\\
 \caption{Graphical comparison of results using the Padova models and noisy data, ratio S/N=5.
Figure (a) from top to bottom
 and shifted by a constant to aid visualization: noisy test spectrum,
 spectrum recovered using active learning and the original data and spectrum recovered
using active learning combining predictions. Figures (b) and (c) show the relative difference
 between test spectrum and predicted spectra in the same listed order.}
  \label{f:ensPado}
\end{figure}

These results were the most accurate ones, even with high levels
of noise, they are presented in
Figures \ref{f:histAgeEnsGranMuchoRuido} to
\ref{f:histRedEnsGranMuchoRuido}. These figures show a marginal
improvement with respect to those of Figures
\ref{f:histAgeGranMuchoRuido} to \ref{f:histRedGranMuchoRuido}. A
graphical comparison is presented in Figures~\ref{f:ensGran} and
\ref{f:ensPado}. 
It may be argued that the inclusion of constant Gaussian noise to
the synthetic spectrum will produce a low S/N in those regions with lower
signal and this will preferentially affect the
Lick indices. While some of this is present for the deepest features, 
it cannot be a major effect for the large majority of the Lick indices 
where the flux in the control band only changes by less than 20 percent in 
average with respect to the side bands.
The improvement in the concentration of results is clearly illustrated in
Table 4 where the central
bin frequency increases substantially by the inclusion of prior knowledge.
In general, the results obtained from both, the
Padova and the Granada models, support the conclusion that the
best method when dealing with low S/N data seems to be the
combination of ensemble and Prior knowledge.

Results are similar for both sets of theoretical models. However,
we can point some interesting differences in the experimental
results. For instance, when using noiseless data the method
achieves slightly higher prediction accuracies for the Granada
models. Another difference mentioned previously is that for the
Granada models prior knowledge does not seem to be very useful by
itself. Although, the only method that for this models yields
better results than active learning with original data, is the
combination of the three predictions: original data plus the two
methods for using prior knowledge. In contrast, for the Padova
models both methods for using prior knowledge improved prediction
accuracy. It should be emphasized that these differences between
the models are not significative, and we do not consider they can
be thought of as evidence of the correctness of the models, but
as the relation the spectrum has to the set of parameters on each
model. Results on both models show that the ensemble of
classifiers combining different forms of incorporating prior
knowledge is the best alternative, specially when the data have 
high levels of noise.

\begin{table*}
\centering
\begin{tabular}{|c|c|ccccccc|}
\hline
 &  & \multicolumn{7}{|c|}{Bin Centres/Frequencies}\\
\hline Algorithm & Age & -3 & -2 & -1 & 0 & 1 & 2 & 3\\
\hline \multicolumn{1}{|l|}{LWLR ensamble} &
\multicolumn{1}{|l|}{Intermediate Population} & 0 & 15 &28 &88
&15& 4&
0\\
 & \multicolumn{1}{|l|}{Old Population}& 8&10&15& 46& 31& 24& 16\\ \hline
\multicolumn{1}{|l|}{Active Learning} &
\multicolumn{1}{|l|}{Intermediate Population} & 7 &13 &22 &95 &11&
2&
0\\
& \multicolumn{1}{|l|}{Old Population} & 7& 6& 18&42&27&26&24\\
\hline \multicolumn{1}{|l|}{Ensemble \& Prior knowledge}&
\multicolumn{1}{|l|}{Intermediate Population} & 4 &15 &18&
99& 10& 4 &0\\
& \multicolumn{1}{|l|}{Old Populations} & 4 & 7& 18&55&26& 26&14\\

\hline
\end{tabular}
\caption{Frequency Table for prediction of ages using the Padova
models and an S/N=5. \label{t:AgePado}}
\end{table*}

\begin{table*}
\centering
\begin{tabular}{|c|c|ccccccccccccc|}
\hline
 &  & \multicolumn{13}{|c|}{Bin Centres/Frequencies}\\
\hline Algorithm & Age & -0.9 & -0.75 & -0.6 & -0.45 & -0.3 & -0.15 & 0 & 0.15&0.3& 0.45& 0.6&0.75&0.9 \\
\hline \multicolumn{1}{|l|}{LWLR ensamble} & \multicolumn{1}{|l|}{Young Population}  & 1    & 4   & 10   & 13   & 10   & 13   & 27   & 23   & 30   & 15    & 3    & 0   &  1\\
 & \multicolumn{1}{|l|}{Intermediate Population} & 1    & 1    & 4   &  7  &  16   & 23   & 27   & 34   & 27     &9    & 1    & 0& 0\\
 & \multicolumn{1}{|l|}{Old Population}& 0    & 0    & 7   & 13   & 13   & 31   & 31  &  22   & 22    & 9    & 2   &  0    & 0\\ \hline
\multicolumn{1}{|l|}{Active Learning} & \multicolumn{1}{|l|}{Young Population} & 0   &  3   &  6   &  7  &  11   & 18   & 35  &  36  &  22   & 10   &  2   &  0   &  0\\
 & \multicolumn{1}{|l|}{Intermediate Population} & 1    & 0    & 8   &  8  &  11   & 21   & 36   & 36    &22    & 7    & 0& 0& 0\\
& \multicolumn{1}{|l|}{Old Population} & 0    & 0    & 5   & 12 &
11   & 22   & 45   & 27 &19& 8    & 1    & 0    & 0\\ \hline
\multicolumn{1}{|l|}{Ensemble \& Prior
knowledge}& \multicolumn{1}{|l|}{Young Population}& 0    & 0   &  3   &  6 &   19   & 28  &  49  &  28  &  14   &  2   &  1   &  0   &  0\\
& \multicolumn{1}{|l|}{Intermediate Populations} & 1  &   0 &    2
&    6 &    7 & 33 &
51 &30 &   15&     5 &    0&     0 &    0\\
& \multicolumn{1}{|l|}{Old Populations} & 0  &   0 &    1 &    4  &   6  &  23 &   63  &  33 &   17 &    3 &    0 &    0 &    0\\
\hline
\end{tabular}
\caption{Frequency Table for prediction of relative contributions
using the Padova models and an S/N=5. \label{t:ContPado}}
\end{table*}

\begin{table*}
\centering
\begin{tabular}{||c|c||ccccccccc||}
\hline
 &  & \multicolumn{9}{||c||}{Bin Centres/Frequencies}\\
\hline Algorithm & Age & -0.0006& -0.00045&
-0.0003&-0.00015&0& 0.00015&0.0003&0.00045&0.0006\\
\hline \multicolumn{1}{|l|}{LWLR ensamble} & \multicolumn{1}{|l||}{Young Population}  & 12  &   8  &  22  &  24  &  23  &  20  &  16  & 16   &  9\\
 & \multicolumn{1}{|l||}{Intermediate Population} & 11  &  19   & 12 &   28   & 21 &   19   & 18  &  13  &   9\\
 & \multicolumn{1}{|l||}{Old Population}& 14 &   19  &  19  &  19  &  16  &  18  &  14    &22    & 9\\ \hline
\multicolumn{1}{|l||}{Active Learning} & \multicolumn{1}{l}{Young Population}&  3   &  9  &  21  &  23  &  24  &  23  &  28  &  15   &  4\\
 & \multicolumn{1}{|l||}{Intermediate Population} & 7  &  14  &  21  &  27 &   27 &   21  &  17  &  11  &   5\\
& \multicolumn{1}{|l||}{Old Population} & 6 &   11&    19  &  28&    37&    23&    12 &    9& 5\\
\hline \multicolumn{1}{|l||}{Ensemble \& Prior
knowledge}& \multicolumn{1}{|l||}{Young Population}& 1  &  10  &  14  &  23  &  47  &  29  &  19   &  6   &  1\\
& \multicolumn{1}{|l||}{Intermediate Populations} & 0   &  5  &  29  &  19  &  43  &  29    &16    & 6    & 3\\
& \multicolumn{1}{|l||}{Old Populations} & 2   &  7 &   16  &  30  &  39  &  26   & 24   &  5  &   1\\
\hline
\end{tabular}
\caption{Frequency Table for prediction of reddening parameters
using the Padova models and an S/N=5. \label{t:RedPado}}
\end{table*}

\section{Conclusions}

We presented in this work an optimization algorithm that can
estimate with high accuracy:  age distributions and mixtures plus
the reddening of stellar population in galaxies. The algorithm
achieves convergence by iteratively creating new data points that
lie in the vicinity of the query point.

Our experimental results using two sets of theoretical
models and different levels of noise, show that
even with low quality (S/N=5) data the algorithm does  a good
estimate of the population ages, proportions and reddening.
For our method in about 80\% of
the cases the error in the age determination is equal or less than
one age step. In general, the results obtained from both the
Padova and the Granada models support the conclusion that the best
method when dealing with low S/N data seems to be the combination
of an ensemble and prior knowledge. Another important feature
of this method is its high speed, it takes $\sim$10 seconds in a
normal PC to estimate
the parameters of a single 20,000 pixel spectrum. This represents
a great advantage over other more conventional methods proposed
for this problem, which may take up to a couple of hours to find the
solution for such a spectrum.

We will continue our efforts to improve parameter estimation of
stellar populations. In forthcoming papers we experiment with
models of different metallicities, by adapting this method
successfully  to this problem. Also, we explore different methods
for exploiting prior knowledge and apply them to large spectral
databases (e.g.~SDSS).

Based on this experimental evaluation we conclude that this method
can be applied with similar success to ``real" galaxies, reducing the
computational cost and thus providing the capability of analyzing large
quantities of astronomical spectroscopic data.

\section*{Acknowledgments}
We would like to acknowledge financial support from CONACYT,
the Mexican Research Council, through research grants \#~166934,
\#~32186-E and \#~40018-A-1. ET and RJT are grateful for the
hospitality of the IoA, Cambridge, where part of this work was
accomplished. We are indebted to Sandro Bressan and Miguel
Cervi\~no and their colleagues, for generously providing their
high resolution models before publication and for very
constructive discussions. We also enjoyed discussing this work with 
Roberto Cid Fernandes.

\bsp

\label{lastpage}

\end{document}